\documentclass[pra, twocolumn,amsmath,amssymb,superscriptaddress]{revtex4-2}

\usepackage{amsfonts}
\usepackage{physics}
\usepackage{graphicx}
\usepackage{natbib}
\usepackage{tikz}
\usepackage{soul}
\usepackage{ragged2e}
\usepackage{blindtext}
\usepackage{mathtools}
\usepackage{soul}

\usepackage{amsmath}
\usepackage{xcolor}

\usepackage{mathptmx}           
\usepackage{graphicx}           
\usepackage{subcaption}
\usepackage{url}                

\usepackage{textcomp}
\usepackage{float}

\newcommand{\overbar}[1]{\mkern 1.5mu\overline{\mkern-1.5mu#1\mkern-1.5mu}\mkern 1.5mu}

\newcommand{\bea}{\begin{eqnarray}}
\newcommand{\eea}{\end{eqnarray}}
\newcommand{\be}{\begin{equation}}
\newcommand{\ee}{\end{equation}}

\newcommand{\svloss}{\left<\sigma_{\rm{loss}} \ v\right>}

\newcommand{\kb}{k_{\rm{B}}}

\newcommand{\Ud}{U_{\rm{d}}}
\newcommand{\Udsix}{U_{\rm{d},6}}

\newcommand{\svtot}{\left<\sigma_{\rm{tot}} \  v\right>}

\newcommand{\svtotSC}{\left<\sigma_{\rm{tot}} \  v\right>_{\mathrm{SC}}}

\newcommand{\svtotMED}{\left<\sigma_{\rm{tot}} \  v\right>_{\mathrm{Med}}}

\newcommand{\svtotKT}{\left<\sigma_{\rm{tot}} \  v\right>_{\mathrm{KT}}}

\newcommand{\gtot}{\Gamma_{\rm{tot}}}
\newcommand{\gloss}{\Gamma_{\rm{loss}}}

\newcommand{\vp}{v_{\rm{p}}}

\newcommand{\vg}{v_{\rm{G}}}

\newcommand{\pqdu}{p_{\rm{QDU}}}
\newcommand{\thetamin}{\theta_{\rm{min}}}

\newcommand{\svtotCsix}{\left<\sigma_{\rm{tot}} v \right>_{\rm{C6}}}

\newcommand{\uf}{p_{\mathrm{QDU}}}
\newcommand{\vmaxone}{v_{\rm{max}}}
\newcommand{\vdvmax}{v_{vdv}^{\rm{max}}}

\usepackage{xspace}

\begin{document}

\title{Revising the Universality Hypothesis for Room-Temperature Collisions}

\author{James L. Booth} \email{jbooth4@my.bcit.ca}
\affiliation{Physics Department, British Columbia Institute of Technology, 3700 Willingdon Avenue, Burnaby, B.C. V5G 3H2, Canada}

\author{Kirk W. Madison} \email{madison@phas.ubc.ca}
\affiliation{Department of Physics \& Astronomy, University of British Columbia,\\
6224 Agricultural Road, Vancouver, B.C., V6T 1Z1, Canada}

\date{\today}

\begin{abstract}
\noindent{Atoms constitute promising quantum sensors for a variety of scenarios including vacuum metrology. Key to this application is knowledge of the collision rate coefficient of the sensor atom with the particles being detected. Prior work demonstrated that, for room-temperature collisions, the total collision rate coefficient and the trap depth dependence of the sensor atom loss rate from shallow traps are both universal, independent of the interaction potential at short range.  It was also shown that measurements of the energy transferred to the sensor atom by the collision can be used to estimate the total collision rate coefficient.
However, discrepancies found when comparing the results of this and other methods of deducing the rate coefficient call into question its accuracy.  Here the universality hypothesis is re-examined and an important correction is presented. 
We find that measurements of the post-collision recoil energy of sensor atoms held in shallow magnetic traps only provide information about the interaction potential at the very largest inter-atomic distances (e.g.~the value of $C_6$ for a leading order term of $C_6/r^6$).  As other non-negligible terms exist at medium and long ranges, the total collision rate coefficient, even if universal, can differ from that computed solely from the value of $C_6$. By incorporating these other long-range terms into a simple semi-classical (SC) calculation, we find the SC prediction matches that of full, multi-channel, quantum mechanical scattering calculations using the complete potential.  This work resolves the discrepancies, demonstrates the simplicity of estimating the rate coefficients for universal collision partners, and provides guidance for using atoms as a self-calibrating primary quantum pressure standard.}

\end{abstract}


\maketitle

\section{Introduction}

Laser cooled atoms have been proposed \cite{booth2011,madison2012,PhysRevA.85.033420} and investigated \cite{yuan2013,Rev.Sci.Ins.093108.2015,makhalov2016,makhalov2017,Julia2017,Julia2018,Eckel_2018,xiang2018,Booth2019,Shen_2020,Shen_2021,BARKER2021100229,doi:10.1116/5.0095011,zhang2022,Ehinger2022,SUN2024113079} as a sensor for particles in vacuum.  Specifically, the ambient gas number density, $\rho_N$, can be determined from measurements of the total collision rate of a stationary sensor atom with the ambient gas particles
\be
\gtot = \rho_N \svtot(T)
\label{eq:mastergamma}
\ee
where $\svtot$ is referred to as the total collision rate coefficient and depends on the ambient gas temperature ($T$), $\sigma_{\rm{tot}}$ is the velocity-dependent cross section for the ambient gas and sensor particle collision, $v$ is the relative collision velocity, and the brackets denote an average over the Maxwell Boltzmann distribution of velocities for the ambient gas.

In practice, the sensor atoms are confined in a shallow trap, and the total collision rate is inferred from the observed trap loss rate,
\bea
\Gamma_{\rm{loss}} = \rho_N\svloss(T,U),
\label{eq:mastergammaloss}
\eea
where $\svloss$ is the thermally-averaged loss rate coefficient. This coefficient depends on both the ambient gas temperature, $T$, and on the depth of the trap, $U$, confining the sensor atoms. In order for sensor atom loss to occur, a collision must transmit sufficient energy to the trapped atom that its energy exceeds the trap depth. In the limit $U \rightarrow 0$, the trap loss rate becomes the total collision rate, $\gloss \rightarrow \gtot$.

Deducing $\rho_N$ from Eq.~\ref{eq:mastergamma} or Eq.~\ref{eq:mastergammaloss} requires accurate knowledge of the rate coefficients.  These coefficients can be inferred from measurements of the sensor atom loss rate at various trap depths, $\Gamma_{\rm{loss}}$, when exposed to a background gas of known density, $\rho_N$ \cite{Barker2023,Eckel2023}.  Unfortunately, this approach is limited to inert gas species compatible with the operation of existing orifice flow pressure standards (OFS).  Alternatively, if the interaction potential between the sensor atom and the species of interest is known, the coefficient can be obtained from full quantum mechanical scattering calculations (FQMS)
\cite{PhysRevA.99.042704,PhysRevA.101.012702,PhysRevA.105.029902,PhysRevA.105.039903,Klos:2023}. 

A third method is to use the
so-called collision universality law to estimate the coefficient from measurements of the sensor atom collision recoil energy distribution \cite{Booth2019,Madison2018,Shen_2020,Shen_2021}. 
This approach was used to estimate the coefficients for $^{87}$Rb sensor atoms with natural abundance H$_2$, He, N$_2$, Ar, CO$_2$, Kr, and Xe gases in Refs.~\cite{Shen_2020,Shen_2021} and for natural abundance Rb gas collisions in Ref.~\cite{PhysRevA.106.052812}. Following that work, results from both theory calculations (FQMS) and orifice flow experiments (OFS) revealed differences in the values, at the level of a few percent, for the $^{87}$Rb-X coefficients given X=N$_2$, Ar, Kr, and Xe gas particles as shown in Table~\ref{tab:Rb_summary}.  For X=H$_2$ gas, the FQMS result differed by more than 25\% from the experimentally determined value.  This motivated work that confirmed this larger discrepancy for $^{87}$Rb-H$_2$ is due to the previously anticipated breakdown of the universality prescription for determining the coefficient between light collision partners \cite{Shen_2023}.  

The main goal of this work is to re-examine the original postulates of the collision universality law in an effort to understand the discrepancies
observed in the $\svtot$ values deduced for heavy collision partners using the universality prescription versus those found by FQMS reported in Ref.~\cite{Klos:2023} and the OFS measurements reported in Refs.~\cite{Barker2023,Eckel2023}.  

The original universality hypothesis included three conjectures,
\begin{enumerate}
    \item The total collision rate coefficient, $\svtot$, is universal, determined solely by the long-range portion of the inter-species interaction potential. Namely, $\svtot$ is independent of the details of the interaction potential at short range.
    \item The trap loss rate coefficient is also universal, where $\rho_N \svloss (U) $  is the rate of collisions that induce loss from a trap of depth $U$.
    \item The variation of the loss rate coefficient with trap depth provides information about the total collision rate coefficient, $\svtot$.
\end{enumerate}

We show here, with data and computations, that first two postulates are sound but the third postulate needs revision.  Specifically, we find that the sensor-atom collision-induced recoil energy distribution near zero energy, encoded in the loss rate versus trap depth from shallow traps ($U/\kb < 20$~mK), provides information about the leading order term, $-C_6/r^6$, of the interaction potential instead of the total collision rate coefficient.  Although $\svtot$ is dominated by the $C_6$ term and, for heavy collision partners, is insensitive to the shape of the interaction potential at short range, there are additional long range interactions, namely the $-C_8/r^8$ and $-C_{10}/r^{10}$ terms, that contribute to the total collision rate coefficient and make it distinct from that deduced using the universality prescription \cite{Booth2019,Madison2018,Shen_2020,Shen_2021}. By including these 
long-range interaction terms in a semi-classical analysis that provides a prediction of $\svtot$ in the absence of any short range repulsive barrier, we find the prediction for heavy collision partners to be in excellent agreement with the full quantum scattering calculations reported in Ref.~\cite{Klos:2023} at ambient temperature 21~C (294 K), differing by less than the stated uncertainties.





\begin{table}[h]
\centering
\begin{tabular}{llll}
\hline\hline
System  & \multicolumn{3}{c}{$\svtot$ ($10^{-15}$m$^3$/s)} \\
 & \multicolumn{1}{c}{QDU \cite{Shen_2021}} & \multicolumn{1}{c}{FQMS \cite{Klos:2023}} &\multicolumn{1}{c}{OFS
  \cite{Barker2023}}  \\
\hline 
 $^{87}$Rb-H$_2$ &  5.12(15) &  3.9(1) &\multicolumn{1}{c}{---} \\
 $^{87}$Rb-He &  2.41(14) & 2.37(3) & 2.34(6) \\
 $^{87}$Rb-Ne &  \multicolumn{1}{c}{---}  & 2.0(2) & 2.21(5) \\
 $^{87}$Rb-N$_2$ & 3.14(5) & 3.45(6) &3.56(8) \\
 $^{87}$Rb-Ar &  2.79(5) & 3.035(7) & 3.31(5) \\
 $^{87}$Rb-CO$_2$ & 2.84(6) & \multicolumn{1}{c}{---} & \multicolumn{1}{c}{---} \\
 $^{87}$Rb-Kr &  \multicolumn{1}{c}{---} & 2.79(1) & 2.79(4) \\
 $^{87}$Rb-Xe &  2.75(4) & 2.88(1) & 2.95(7) \\
\hline
\end{tabular}
\caption{\justifying{
Total collision rate coefficients, $\svtot$, for $^{87}$Rb when exposed to natural abundance versions of various species at a temperature of T=294 K as determined by the universal law of quantum diffractive collisions (QDU), FQMS calculations, and comparison with an orifice flow standard (OFS). All uncertainties are one-standard deviation $k=1$ (statistical) uncertainties.}}
\label{tab:Rb_summary}
\end{table}

This work is organized as follows: Section~\ref{TOS} reviews the theoretical framework for atom-atom elastic scattering and the original universality ansatz presented in Refs.~\cite{Booth2019, Shen_2020}. Section~\ref{Basis} demonstrates the physical origins of universality.  It is shown that the influence of the potential shape at short range on the total collision rate coefficient, $\svtot$, is minimized due to the averaging of the velocity-dependent collision cross section over the room-temperature Maxwell-Boltzmann (MB) distribution of incident speeds.  We show that above a critical velocity, the collision cross section transitions from a trend dictated by the long range interaction potential to a monotonic variation determined by the short range interaction potential.  If the MB distribution contains no significant weight above this velocity, the total rate coefficient is universal.  This furnishes a heuristic criterion for universality and explains its breakdown at high temperatures and for background species with small mass. 
Section~\ref{TNG} analyzes the effects of including the $C_8$ and $C_{10}$ long range interaction terms omitted in the original analysis.  Using a purely long-range interaction potential of the form $V(r) = - C_6/r^6 - C_8/r^8 - C_{10}/r^{10}$, the total collision rate coefficients computed using a semi-classical (SC) description of the elastic scattering phase shifts are found to be in excellent agreement with FQMS computations using the interaction potentials published by K{\l}os and Tiesinga \cite{Klos:2023}.  We achieve this by extending the Jeffreys-Born approximation \cite{child1996molecular}, and the agreement of the FQMS and SC values of $\svtot$ reinforces the universality hypothesis and the simplicity of estimating the rate coefficients for universal collision partners. Section~\ref{Refine} uses FQMS computations for $^7$Li-X and $^{87}$Rb-X (X = H$_2$, He, Ne, N$_2$, Ar, Kr, and Xe) collisions based on the interaction potentials published by K{\l}os and Tiesinga \cite{Klos:2023} to re-examine and redefine the relationship between $\svloss$ and $\svtot$.  Specifically, we find that the variation of the trap loss rate with trap depth provides the quantum diffractive collision energy scale set by the leading order interaction potential term, $-C_6/r^6$.  The consequence of this is that the self-calibration method underestimates the total rate coefficient by up to 10\% for a universal collision partner.  Finally, the appendix provides an analysis of the origin and character of the glory undulations in the cross section as a function of velocity, serving as a supplement to the discussion in the main text.  The appendix also provides an analysis of the Jeffreys Born approximations used in this work.

We believe this work explains the observed discrepancies, reinforces the conjecture that room-temperature collisions are universal, insensitive to the short range interaction potential shape, demonstrates the simplicity of accurately estimating the rate coefficients for universal collision partners using a semi-classical approximation, and provides guidance for using atoms as a self-calibrating primary pressure standard.

\section{Review of the Universality Hypothesis}\label{TOS}


In this section, we review the quantum mechanical formulation of the total collision rate coefficient, $\svtot$, and the trap loss rate coefficient, $\svloss (U)$.  We also present calculations of these coefficients using a Lennard-Jones interaction potential to illustrate the original universality ansatz presented in Refs.~\cite{Booth2019, Shen_2020}.

We consider a collision between a stationary trapped atom of mass, $m_t$, and a background gas particle of mass $m_b$, whose speed is selected from the Maxwell-Boltzmann (MB) distribution of the particles at ambient temperature, $T$. The collision is analyzed in the center of mass (COM) frame where a reduced mass particle, 
\be
\mu = \frac{m_t m_b}{m_t + m_b}
\label{eq:mu}
\ee
scatters from the interaction potential, V(r), or from a potential energy surface, for lower symmetry collisions. In what follows, we will assume that the interaction potential is spherically symmetric.  Although there exist anisotropic interactions arising from, for example, magnetic dipole-dipole interactions, these are exceedingly weak compared to the van der Waals interaction potential and are neglected in the following analysis for atom-atom collisions.  Below, we discuss the consequence of this spherical symmetry assumption on the results of our rate coefficient calculations.  We will also assume that each collision liberates a single trapped atom, valid for the exceedingly low density of the trapped ensembles employed (for justification, see for example Ref.~\cite{PhysRevA.109.032818}).

For elastic scattering, the magnitude of the reduced mass particle's momenta before ($\vec{p} = \hbar \vec{k}$) and after the collision ($\vec{p}^{\prime} = \hbar \vec{k}^{\prime}$) are equal.  Only the direction of the momentum of the reduced mass particle is altered by the collision, rotating an amount $\theta$. Here $k = \mu v/\hbar$ is the magnitude of the wavevector describing the momentum of the incoming and outgoing reduced-mass particle, and $v$ is the relative speed of the colliding particles. From the kinematics of elastic collisions, one can deduce that the energy transferred to the (initially stationary) trapped particle in the laboratory frame is,
\be
\Delta E_{t} = \frac{\mu^2 v^2}{m_t}\left(1-\cos\theta\right).
\label{eq:DeltaEtrapped}
\ee
In order to liberate a sensor atom from the trap, this energy must be greater than or equal to the trap depth, $U$ (for a stationary trapped atom). Rearranging Eq.~\ref{eq:DeltaEtrapped} and setting $\Delta E_t = U$ one finds the corresponding minimum scattering angle required for trap loss is
\be
\cos\thetamin =  1 - \frac{m_t U}{\mu^2 v^2},
\label{eq:thetamin}
\ee
which is explicitly dependent on the trap depth, $U$.


In the asymptotic region, far from the scattering center, the wave-function, in the COM frame, is the superposition of an incident plane wave and the scattered spherical wave, $\psi({\bf r}) \xrightarrow[]{{\bf r} \rightarrow \infty} e^{ikz} + f(k,\theta) \frac{e^{ikr}}{r}$, where $f(k,\theta)$ quantifies the amplitude of scattering into the angle $\theta$ given an incident momentum $\hbar k$. The dependence of $f$ on $\phi$ is absent here since the scattering process does not change the azimuthal angle for a spherically symmetric potential.
This scattering amplitude can be expressed as a sum of Legendre polynomials, $P_L\left(\cos\theta\right)$, weighted by the transition matrix elements $T_L(k)$, encoding the scattering amplitude into a particular ``partial wave" with angular momentum $\hbar L$ as
\be
f(k,\theta) = \frac{1}{k}\sum_L \left(2L+1\right) T_L(k) P_L\left(\cos\theta\right).
\label{eq:fktheta}
\ee

The total elastic collision cross-section for a scattering event with wavevector $k$ is then the sum of the scattering probability over all angles, 
\bea
\sigma_{\rm{tot}}\left(k\right) 
&=& \int d\Omega \left| f(k, \theta)\right|^2 \nonumber \\
&=& \frac{1}{k^2}\int_{0}^{2\pi} d\phi \int_{0}^{\pi}\sin\theta d\theta \left[\sum_L \left(2L+1\right) T^*_L(k) P_L\left(\cos\theta\right) \right] \nonumber \\
& & \ \ \ \ \ \ \ \ \ \ \ \ \ \ \ \ \ \ \ \ \ \ \ \times \left[\sum_{L^{\prime}} \left(2L^{\prime}+1\right) T_{L^{\prime}}(k) P_{L^{\prime}}\left(\cos\theta\right) \right].
\label{eq:sigmatotk}
\eea

The thermally averaged, total collision coefficient, $\svtot$, is computed by averaging $\sigma_{\rm{tot}}(k)$ over the distribution of relative velocities in the center-of-mass frame.  To compute this average, we approximate the relative speed distribution of the collision partners using the lab frame Maxwell-Boltzmann (MB) speed distribution, $d(v)$, of the background gas particles at room temperature.  This approach is exact in the limit that the trapped sensor particle is stationary, and,  as shown in Ref.~\cite{PhysRevA.109.032818}, this approximation is justified here because the energy of the sensor atoms in the trap is on the order of $10^6$ times smaller than the energy of the background gas particles.
\bea
\nonumber
\svtot & = & \int_{0}^{\infty} d(v) dv \left( 2\pi \cdot v \int_{0}^{\pi} \sin\theta d\theta \left|f\left(k, \theta\right)\right|^2 \right)\\
& = & \left< 2\pi \cdot v \int_{0}^{\pi} \sin\theta d\theta \left|f\left(k, \theta\right)\right|^2 \right>
\label{eq:svtot}
\eea
where the speed distribution is
\bea
d(v) dv 
& =& \frac{4}{\sqrt{\pi}}\left(\frac{v}{v_p}\right)^2 e^{-\left(\frac{v}{v_p}\right)^2} \frac{dv}{v_p}.
\label{eq:MBvdist}
\eea
Here $v_p = \sqrt{2 k_B T/m_b}$ is the peak speed of the distribution.

Trap loss only occurs for $\theta \in [\theta_{\rm{min}},\pi]$, leading to the thermally averaged trap loss rate coefficient,
\be
\svloss = \left< 2\pi \cdot v \int_{\thetamin}^{\pi} \sin\theta d\theta \left| f\left( k, \theta \right) \right|^2 \right>.
\label{eq:svloss}
\ee
Here $\thetamin$ is given by Eq.~\ref{eq:thetamin} and is dependent on both the relative collision velocity, $v$, and the sensor atom trap depth, $U$. Without loss of generality, one can re-write Eq.~\ref{eq:svloss} as,
\begin{eqnarray}
\svloss &=& \left< 2\pi \cdot v\int_{0}^{\pi} \sin\theta d\theta \left|f\left(k, \theta\right)\right|^2 \right> \nonumber \\
& &- \left< 2\pi\cdot v \int_{0}^{\thetamin(U)} \sin\theta d\theta \left|f\left(k, \theta\right)\right|^2 \right>  \nonumber \\
&=& \svtot \left[ 1 - p(U) \right].
\label{eq:svlosspU}
\end{eqnarray}
The quantity,
\be
p(U) = \frac{\left< 2\pi \cdot v\int_{0}^{\thetamin(U)} \sin\theta d\theta \left|f\left(k, \theta\right)\right|^2 \right>}{\svtot}
\label{eq:pU}
\ee
is the cumulative probability that a collision imparts an energy $\Delta E_t \le U$ to the sensor atom.

Evidence that this cumulative probability distribution is a universal function, insensitive to the details of the interaction potential at short range, was reported in Refs.~\cite{Booth2019, Shen_2020, Shen_2021}.  Specifically, it was asserted that $p(U) = \uf(U/\Ud)$ for any collision pair where $\uf(x)$ is a universal function and $\Ud$ is the so-called quantum diffractive energy:
\be
\Ud = \frac{4\pi \hbar^2}{m_t \bar{\sigma}},
\label{eq:Uddef}
\ee
where 
$\bar{\sigma}~=~\svtot/\vp$ is the effective collision cross-section. 
This is the natural energy scale associated with a collision-induced localization of the sensor atom to a region of size $\sqrt{\bar{\sigma}}\ $ \cite{Booth2019}.  As discussed in Ref.~\cite{Bali1999}, the term ``diffractive collisions" comes from the notion that for hard sphere quantum scattering, where $R$ is the range of the interaction potential, half of the total cross section ($\sigma \simeq 2 \pi R^2$) is accounted for by ``classical" scattering with a geometric cross section of $\pi R^2$ and an additional $\pi R^2$ is due to diffraction \cite{child1996molecular}.



\begin{figure}[ht!]
    \begin{subfigure}[t]{0.48\textwidth}
        \centering
        \includegraphics[width=\linewidth]{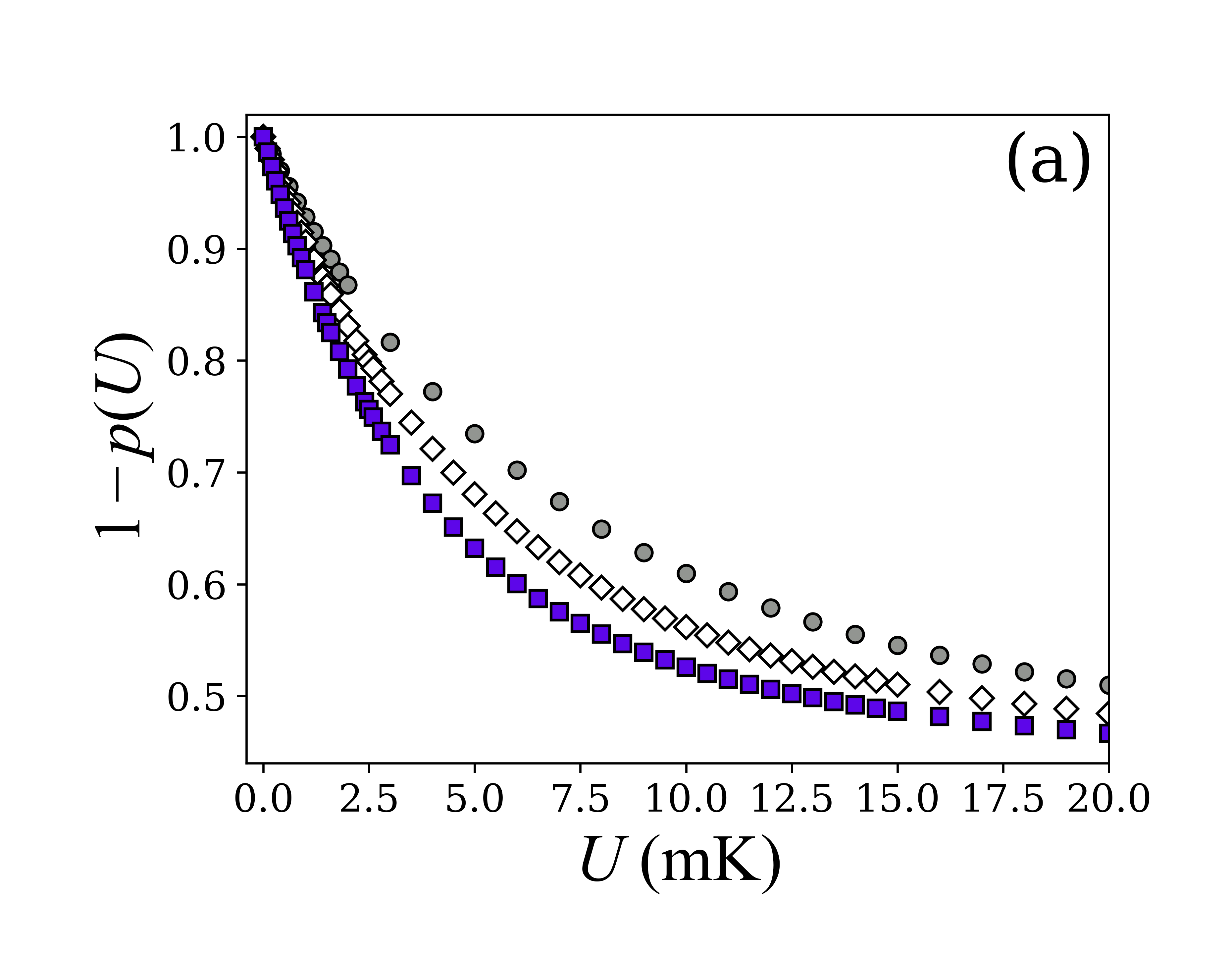} 
        \label{fig:svscaledvsU}
   \end{subfigure}
       \begin{subfigure}[t]{0.48\textwidth}
       \vspace{-5ex}
       \centering
        \includegraphics[width=\linewidth]{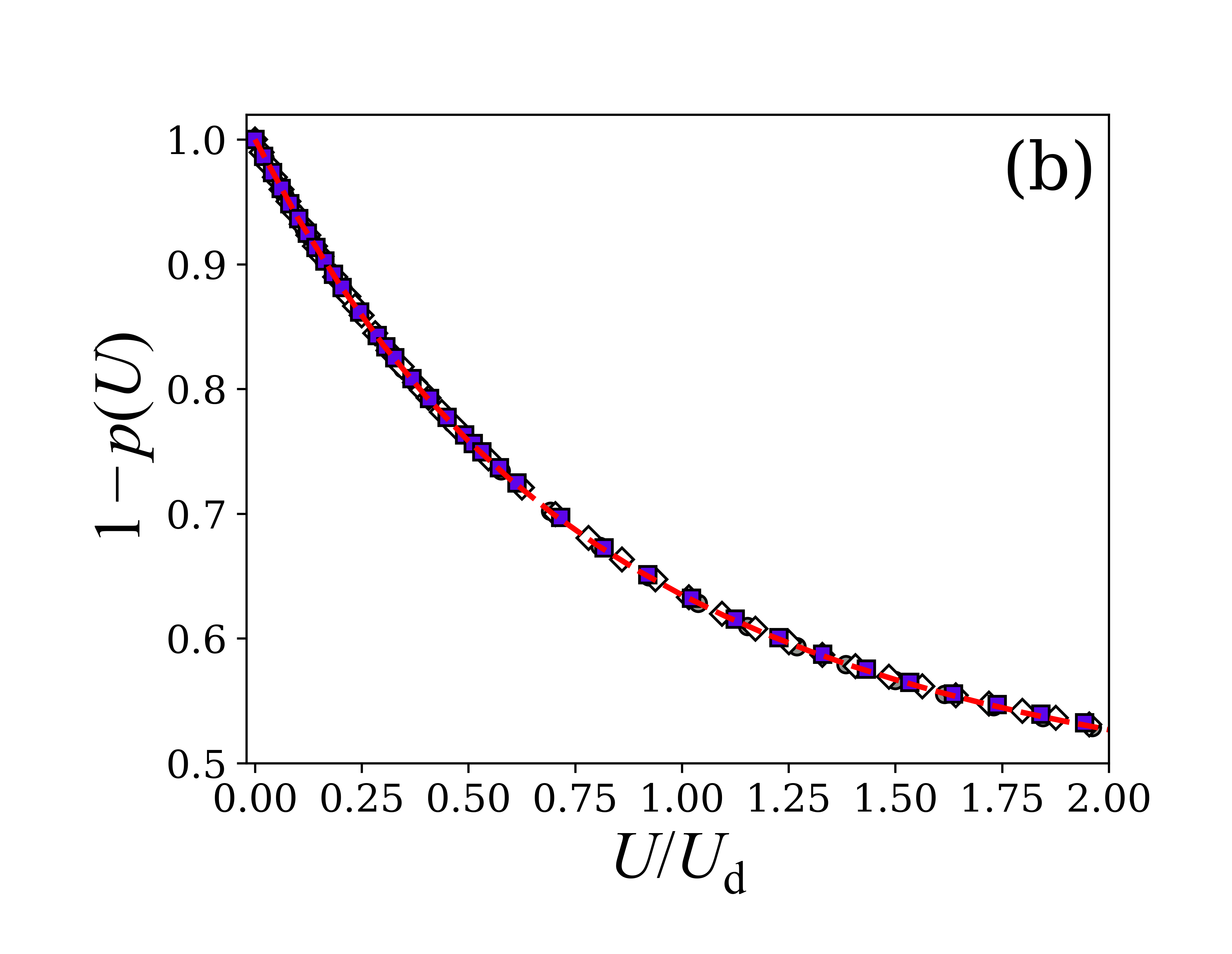} 
        \label{fig:svscaledvsUUd}
   \end{subfigure}
    \caption{\justifying{The ratio of the trap loss rate coefficient to the total collision rate coefficient, $\svloss/\svtot = 1-p(U)$, computed using FQMS calculations for a Lennard Jones potential with the $C_6$ and $D_e$ values from Ref.~\cite{Klos:2023} at T = 294~K.  The results for $^{87}$Rb-Ar (grey circles), $^{87}$Rb-Kr (white diamonds), and $^{87}$Rb-Xe (violet squares) are plotted (a) versus trap depth, $U$, and (b) versus $U/\Ud$. The red dashed curve is the polynomial approximation to the universal function, $1-p_{\rm{QDU}}(U/\Ud)$, defined in Eq.~\ref{eq:pqdu} and Table~\ref{tab:betajs}.}}
    \label{fig:RbArKrXeuniversal}
\end{figure}

In the original investigation into universality \cite{Booth2019, Shen_2020}, FQMS computations were carried out with a Lennard-Jones (LJ) model for the interaction potential,
\bea
V(r) &=& \frac{C_{12}}{r^{12}}  -  \frac{C_{6}}{r^{6}} 
\label{eq:LJPotential}
\eea
where $C_{12} = C_6^2/(4 D_e)$ and $D_e$ is the depth of the interaction potential. For illustration purposes, we show here the results of FQMS calculations for the total collision rate and the trap depth dependent loss rate coefficients at an ambient temperature of 294 K using the LJ interaction model. Figure~\ref{fig:RbArKrXeuniversal}(a) shows the ratio $\svloss(U)/\svtot$ versus trap depth for $^{87}$Rb-Ar, $^{87}$Rb-Kr, and $^{87}$Rb-Xe collision partners, using the $C_6$ and $D_e$ values from \cite{Klos:2023}.  This ratio, per Eq.~\ref{eq:svlosspU}, is equal to $1-p(U)$, and the separation of the individual curves in Fig.~\ref{fig:RbArKrXeuniversal}(a) indicates that $p(U)$ is different for each collision pair. However, when the energy axis is re-scaled by the corresponding value of $\Ud$ in each case (defined in Eq.~\ref{eq:Uddef}) the curves, shown in Fig.~\ref{fig:RbArKrXeuniversal}(b) at low trap depths, collapse to the same function.

This observation motivated the authors of Ref.~\cite{Booth2019} to assert that the $p(U)$ functions are given by
\be
p(U) = \uf \left(\frac{U}{\Ud} \right)
\simeq \sum_j \beta_j \left(\frac{U}{\Ud}\right)^j.
  \label{eq:pqdu}
\ee
where the universal function, $\uf$, is approximated by a sixth order polynomial over the range $U/U_{\rm{d}} \le 2$.  The coefficients, $\beta_j$, listed in Table~\ref{tab:betajs}, are common to all universal collision partners \cite{Booth2019}.

\begin{table}[h]
\centering
\begin{tabular}{||c|l||}
\hline
 Coefficient &  \phantom{---}Value   \\
\hline
$\beta_1$ & \phantom{ -}0.673(7) \\
$\beta_2$ & \phantom{ }-0.477(3)\\
$\beta_3$ & \phantom{ -}0.228(6)\\
$\beta_4$ & \phantom{ }-0.0703(42)\\
$\beta_5$ & \phantom{ -}0.0123(14)\\
$\beta_6$ & \phantom{ }-0.0009(2)\\
\hline
\end{tabular}
\caption{\justifying{The coefficients, $\beta_j$, of the polynomial approximation to the universal function, $\uf$, in Eq.~\ref{eq:pqdu} as reported in Refs.~\cite{Booth2019, Shen_2020}}}
\label{tab:betajs}
\end{table}


This formulation provides an empirical method for determining $\svtot$ by 
fitting the normalized loss rate measurements,  $\Gamma_{\rm{loss}}(U)/\Gamma_{\rm{loss}}(U=0)$, at different trap depths to the universal expression, Eq.~\ref{eq:pqdu}, extracting $U_{\rm{d}}$ and then deducing $\svtot$ using Eq.~\ref{eq:Uddef}.
This procedure was used in Refs.~\cite{Booth2019,Shen_2020,Shen_2021} to obtain the total collision rate coefficients from the variation of the loss rate with trap depth.  These results are compared, in Table~\ref{tab:Rb_summary}, to those obtained from FQMS and from loss rate measurements at known gas densities using an orifice flow standard (OFS).  The observed discrepancies call into question the accuracy of this empirical method and are the motivation of the present work.

The LJ potential used here and in prior work on this topic does not provide a physically accurate model of the core repulsion between the colliding partners.  In addition, it 
ignores the $C_8$ and $C_{10}$ long-range terms. 
The former inadequacy is somewhat irrelevant for universal collision partners; however, the result of omitting the long-range terms is that the total collision rate coefficient $\svtot$
computed using the LJ potential for a universal collision pair is not the true rate coefficient, which is influenced by all long range terms. Instead, quantum scattering calculations with the LJ potential yield a rate coefficient equal to that obtained for a purely $C_6$ potential of the form $V(r) = - C_6/r^6$.

Because, as we will show, fitting the normalized loss rate variation with trap depth for a universal collision pair provides the $\Udsix$ value, corresponding to $\svtotCsix$, the fit does furnish the same total collision rate coefficient obtained by FQMS calculations for  LJ potential.  Thus, the authors of Ref.~\cite{Booth2019} concluded, incorrectly, that the empirical $\Ud$ values obtained by fitting the normalized loss rate to the universality law are equal to the total collision rate coefficient.

\section{The basis and breakdown of Collision Universality}\label{Basis}

It is well known from experiments with molecular beams that $\sigma_{\rm{tot}}(k)$ (Eq.~\ref{eq:sigmatotk}) exhibits undulatory variations as a function of the relative-velocity about a trend dictated by the long range part of the interaction potential \cite{10.1063/1.1733383,10.1063/1.1733558,child1996molecular}.  These undulations arise from the core-induced, low angular momentum (low $L$) partial waves scattered in the forward direction (referred to as ``glory scattering") which interfere with the high angular momentum (high $L$) partial wave scattering at small angles associated with the long range portion of the interaction potential.  The iconic halo around an observer's shadow created by optical glory scattering in water droplets of a cloud corresponds to backward ($\theta = \pi$) glory scattering (also observed in molecular beam experiments \cite{10.1063/1.465492}). Whereas forward glory scattering at $\theta = 0$, recently observed in molecular beam experiments \cite{ForwardGloryScatteringPaper}, is responsible for creating the undulations in the total cross section.  Changes of the shape of the interaction potential at short range change the {number, the amplitude, and the locations} of the undulatory variations.  Above a critical velocity, referred to here as $\vmaxone$, the glory oscillations cease and $\sigma_{\rm{tot}}(k)$ transitions to following the monotonic trend set by the short range interaction potential.

\begin{figure}
   \centering
    \begin{subfigure}[t]{0.48\textwidth}
        \centering
        \includegraphics[width=\linewidth]{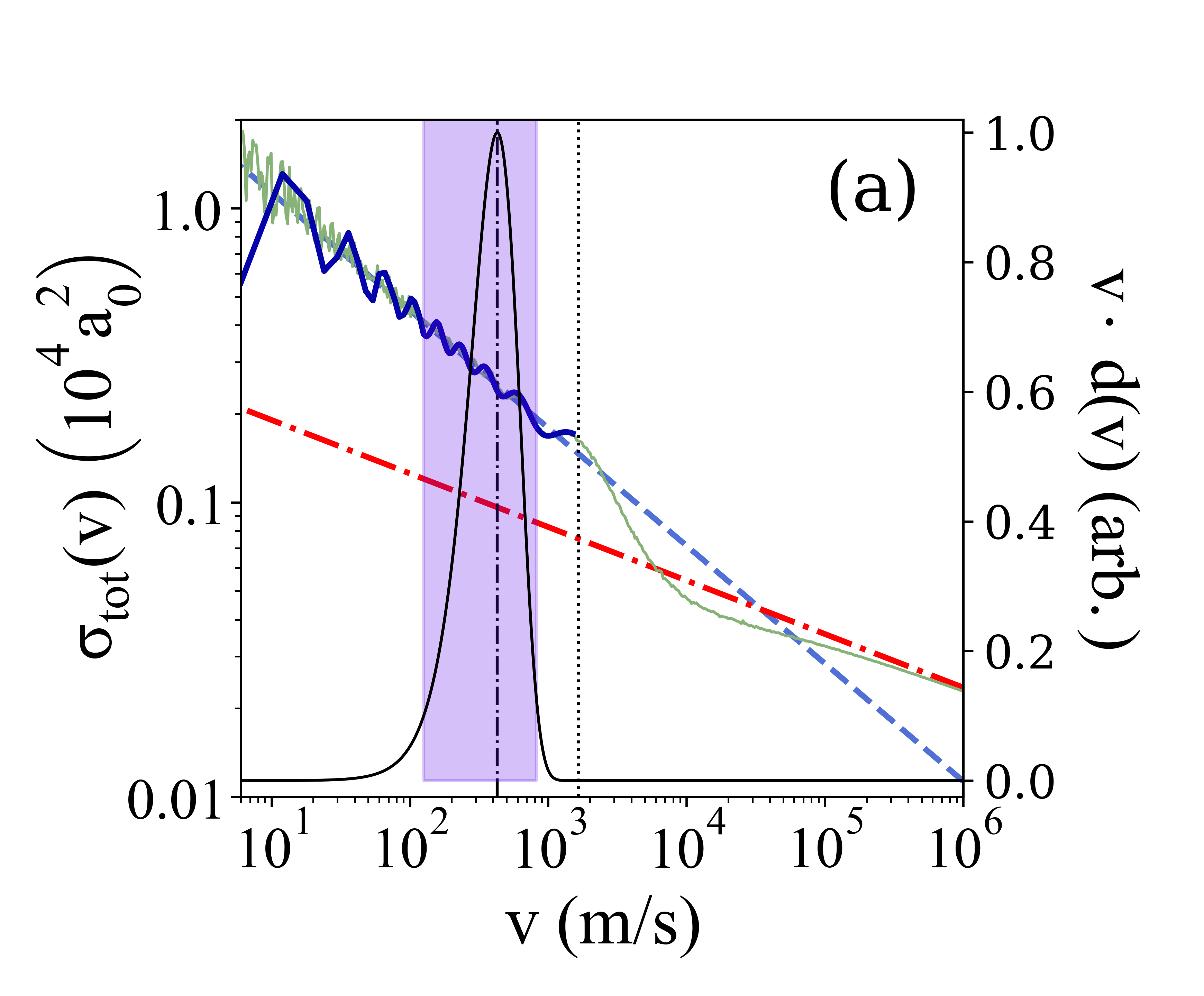} 
         \label{fig:RbArsig}
   \end{subfigure}
   \centering
       \begin{subfigure}[t]{0.48\textwidth}
       \vspace{-5ex}
       \centering
        \includegraphics[width=\linewidth]{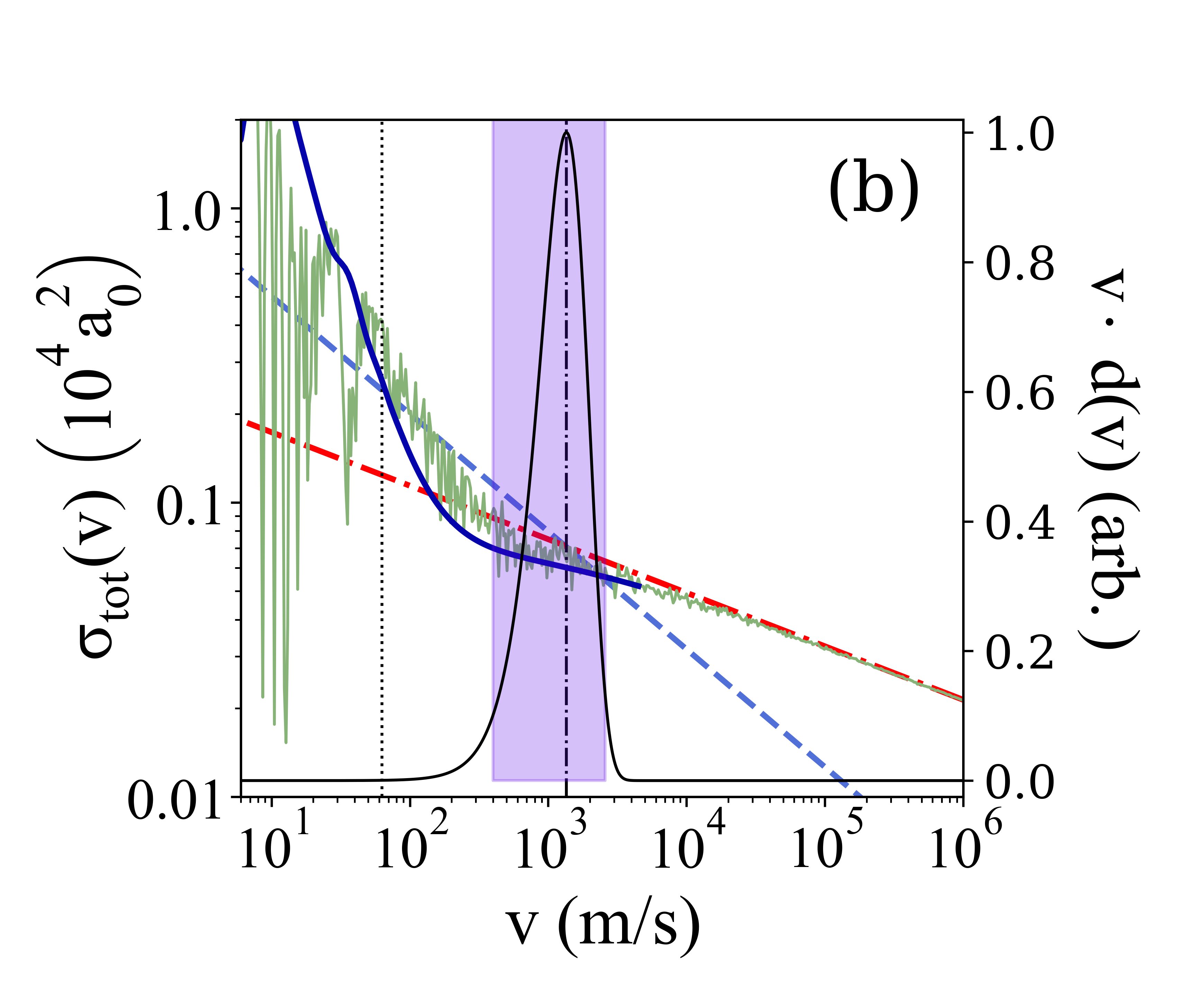} 
        \label{fig:RbHesig}
   \end{subfigure}

    \caption{\justifying{Comparison of the $\sigma_{\rm{tot}}$ versus $v$ for (a) $^{87}$Rb-Ar and (b) $^{87}$Rb-He based on Lennard-Jones (LJ) interaction potentials at T = 294K.  The blue dashed traces are the predictions for a pure long-range $C_6$ potential, Eq.~\ref{eq:stotJBvall}, and the red dot-dashed traces are the pure short-range $C_{12}$ potential predictions, Eq.~\ref{eq:stot12JB}. The thin green traces show $\sigma_{\rm{tot}}(v)$ computed using the LJ potential with the Jeffreys-Born (JB) approximation for the elastic scattering phase shifts, Eq.~\ref{eq:JBetaL6} and \ref{eq:JBeta12}. Note how the green traces follow the pure long range ($C_6$) prediction, then transition towards the pure short-range ($C_{12}$) prediction for large relative collision speeds, $v > \vmaxone$. The FQMS LJ computations are shown as the thick blue traces. The characteristic glory oscillations around the purely $C_6$ predictions are evident for the $^{87}$Rb-Ar results, while these are missing for the low mass collision partner, $^{87}$Rb-He, because the number of bound states is exceedingly low for this latter case (see the appendix). For both figures, the position of the final (largest speed) glory undulation maxmimum is indicated by the vertical dotted black line at $\vmaxone$. Finally, the normalized, velocity-weighted MB distributions, $v\cdot d(v)$ are shown. The shaded region indicates the range over which the normalized $v\cdot d(v) \ge 0.1$. This provides a visual reference for the range of $v$ which contribute the most to the total collision coefficients, $\svtot$. 
    }}
    \label{fig:RbArHeLJcompare}
\end{figure}

To illustrate this behavior Fig.~\ref{fig:RbArHeLJcompare} shows plots of $\sigma_{\rm{tot}}(v~=~\hbar~k/\mu)$ versus $v$, 
for (i) a purely long range $C_6$ potential, of the form $V(r) = - C_6/r^6$ (blue dashed traces), (ii) a purely repulsive $C_{12}$ potential, $V(r) = C_{12}/r^{12}$ (red dot-dashed traces), and (iii) the combination of these two terms to form the Lennard-Jones (LJ) potential where we compute $\sigma_{\rm{tot}}(v)$ using both an approximate solution (solid green trace) and the results of a FQMS calculation (solid blue trace). (Note the FQMS computations were performed over a more limited range of relative collision speeds owing to the increasingly longer computation duration required for convergence of the results at larger $v$ values.)

We determine the scattering cross section by computing the scattering amplitude, Eq.~\ref{eq:fktheta}, where the physics of the interaction is contained in the transition matrix elements
\be
T_L(k) = e^{i\eta_{\rm{L}}(k)}\sin\left(\eta_{\rm{L}}(k)\right).
\label{eq:Tkl}
\ee
The $\eta_{\rm{L}}(k)$ are the momentum-dependent phase shifts for each partial wave.
For a purely long range $C_6$ potential, these phase shifts can be estimated using the Jeffreys-Born (JB) approximation \cite{child1996molecular}, valid for large $L$,
\be
\eta_{\rm{L}}(k, C_6) = \frac{3 \pi}{16} \left(\frac{\mu C_6}{\hbar^2}\right) \frac{k^{4}}{\ell^5},
\label{eq:JBetaL6}
\ee
where $\ell = L + 1/2$. The inclusion of the $1/2$ in this definition of $\eta_{\rm{L}}(k)$ insures that the value remains finite for $L = 0$. In the large $L$ limit the $1/2$ is irrelevant and can be neglected.

Using this in Eq.~\ref{eq:sigmatotk} and approximating the sum over $L$ as an integral over $dL$, we obtain 
\be
\sigma_{\rm{tot}}^{\rm{C6}}(v) \approx 8.0828 \left(\frac{C_6}{\hbar v}\right)^{\frac{2}{5}}
\label{eq:stotJBvall}
\ee
Here the relative speed of the collision, $v = \hbar k/\mu$, is used in place of $k$, and this expression is plotted as the blue dashed traces in Fig.~\ref{fig:RbArHeLJcompare}(a) and (b).  An analogous JB approximation can be found for the phase shifts corresponding to a purely $C_{12}/r^{12}$ repulsive potential:
\be
\eta_{\rm{L}}(k, C_{12}) = -\frac{63 \pi}{512} \left(\frac{\mu C_{12}}{\hbar^2}\right) \frac{k^{10}}{\ell^{11}} = -\frac{63 \pi}{512} \left(\frac{\mu C_6^2}{4 D_e \hbar^2}\right) \frac{k^{10}}{\ell^{11}}.
\label{eq:JBeta12}
\ee
This phase shift yields the velocity dependent cross section,
\bea
\sigma_{\rm{tot}}^{C12}(v) &\approx& 6.5839 \left(\frac{C_{12}}{\hbar v}\right)^{\frac{2}{11}} 
\label{eq:stot12JB}
\eea
corresponding to the red dot-dashed traces on Fig.~\ref{fig:RbArHeLJcompare}(a) and (b).  

The phase shifts for a Lennard-Jones potential can be approximated as the sum of these,
\be
\eta_{\rm{L}}(k, \mathrm{LJ}) = \eta_{\rm{L}}(k, C_{12}) + \eta_{\rm{L}}(k, C_{6}).
\label{eq:JBetaLJ}
\ee
The resulting velocity dependent LJ cross section approximation is shown as the thin green traces on Fig.~\ref{fig:RbArHeLJcompare}(a) and (b).  Also shown in Fig.~\ref{fig:RbArHeLJcompare} are the exact $\sigma_{\rm{tot}}(v)$ found by FQMS calculations (solid blue traces). The agreement is good at large velocities where the JB approximation is valid. (Note the range of $v$ used in the FQMS computation was limited to $0 \le v \le 4\cdot \vp$ due to the rapidly increased computation time needed to achieve convergence for larger $v$ values.) 

As discussed above, the inclusion of a core repulsion in the interaction potential, here a $C_{12}/r^{12}$ term in the LJ potential, modifies the phase shifts to induce an oscillatory behavior in the $\sigma_{\rm{tot}}(v)$ about the purely long range potential prediction, Eq.~\ref{eq:stotJBvall}, at low relative collision speeds.  At large relative collision speeds, the behavior of $\sigma_{\rm{tot}}(v)$ transitions to follow the prediction based on the purely short range part of the potential, Eq.~\ref{eq:stot12JB}.  The transition velocity, defined here as the location of the final, high $v$, glory oscillation maximum, can be found analytically for the LJ potential using the JB approximation. (See Appendix A.)  It is
\bea
\vmaxone &\approx& 
 0.48\cdot \left(\frac{ D_e R_0}{\hbar}\right)
\label{eq:vmax1}
\eea
where 
$R_0 = [C_6/(4 D_e)]^{1/6}$ is the inter-species separation at which the LJ potential is zero (i.e. $V_{\rm{LJ}}(R_0) = 0$). The $\vmaxone$ for the LJ potentials are indicated by the black vertical dotted lines in Fig.~\ref{fig:RbArHeLJcompare}(a) and (b). In short, one may define the long-range ($C_6$) dominated domain as $v \le \vmaxone$.

Also shown in Fig.~\ref{fig:RbArHeLJcompare} is the velocity-weighted MB distribution, $v\cdot d(v)$, (thin black solid trace) for a room-temperature ($T=294$~K) gas of Ar in Fig.~\ref{fig:RbArHeLJcompare}(a) and for He in Fig.~\ref{fig:RbArHeLJcompare}(b). The speed corresponding to the maximum of this normalized distribution, $\vdvmax = \sqrt{3/2}\vp$, is shown as the black vertical dash-dotted line on each figure. The shaded region overlaid on the normalized distributions indicates the range of speeds for which $v\cdot d(v) \ge 0.1$ (i.e. $0.35 \vp \le v \le 2.44 \vp$) as an aid to visualizing the weight and region over which the average $\svtot$ is computed.  The room temperature (T = 294 K) distribution, $v\cdot d(v)$, for $^{87}$Rb-Ar collisions reaches a maximum at a value well below $\vmaxone$ ($\vdvmax < \vmaxone$) and spans several glory oscillations minimizing their influence on the velocity averaged quantities, $\svtot$ and $\svloss$.  This makes these velocity averaged quantities insensitive to the details of the glory undulations and, thus, insensitive to the short range interaction potential shape \cite{Booth2019}. By contrast, one observes that the velocity weighted distribution for $^{87}$Rb-He encompasses the domain where the underlying character of the $\sigma_{\rm{tot}}(v)$ transitions from a purely long-range ($C_6$) to a purely short-range ($C_{12}$) dominated behavior occurring where $\vdvmax > \vmaxone$. Thus, the $^{87}$Rb-He collisions are not universal.

Two properties of the glory oscillations relevant to universality from velocity averaging are \cite{10.1063/1.1733383,10.1063/1.1733558,child1996molecular, PhysRevLett.89.200406} (also see Appendix A),
\begin{enumerate}
\item The amplitude of the glory oscillations decreases as the reduced mass of the colliding partners increases.
\item The number of oscillations per unit speed increases with increased reduced mass of the collision partners.
\end{enumerate}
Thus, for heavier collision partners, one observes low amplitude, high frequency (in speed or wavenumber) glory oscillations leading to an even more complete suppression of the short range influence.

The velocity averaging erasure of the glory scattering effects is the origin of the total collision rate universality for heavy collision partners.  The result is that the $\svtot$ value computed from a given interaction potential is very close to that obtained from a different potential with the same long range behavior but a radically different short range repulsive barrier \cite{Booth2019}.  
This feature is particularly significant to vacuum metrology because it implies that errors or uncertainties in the short-range part of the interaction potential do not propagate to the total collision rate coefficient and thus to the density or pressure inferred from a measurement.

\subsection{Universality breakdown}
Key to universality is the width and position of the relative velocity distribution, $v\cdot d(v)$, compared to the period and the amplitude of the glory undulations associated with the collisions. A hard limit for the breakdown of the universality hypothesis can be defined as when $\vdvmax > \vmaxone$. That is, when the velocity averaging region lies outside the long-range $C_6$ characteristic regime (defined by $v < \vmaxone$) then the short-range character of the collision process becomes dominant. This can occur either because the background gas temperature is too high or because the background species is very light, so the total collision and loss rate coefficients will not be universal, but will depend on the interaction potential at short range.  As seen in Fig.~\ref{fig:RbArHeLJcompare}, significant weight for the $^{87}$Rb-He velocity distribution occurs above $\vmaxone$ (= 62 m/s) and makes the value of $\svtot$ sensitive to the short-range interaction potential shape. 


A second limit for the breakdown of universality occurs if the distribution of relative collision velocities is too narrow compared to the period of the glory undulations, even if $\vdvmax < \vmaxone$. This can happen in a molecular beam experiment or when the background species is extremely cold, rendering the range of velocity averaging small compared to the period of the glory undulations. 

To substantiate this discussion, we compare $\svtot$ values at $T = 294$~K computed by FQMS calculations for a LJ potential with the approximate prediction for a pure long-range $C_6$ potential, 
\be
\svtotCsix \approx 8.49464 \left(\frac{C_6}{\hbar \vp}\right)^{\frac{2}{5}} v_p 
\label{eq:svtot6JB}
\ee 
where $\vp$ is the peak of the MB speed distribution. (See Table~\ref{tab:C6svtot}.)
Although the effects of the short range potential are completely absent in the latter case, the predictions are in close agreement with the FQMS values, and the differences are less than 0.5\% for the heavy collision partners.  However, for collisions with lighter background particles the results are different by up to $\simeq~20$\%, indicating that the $\svtot$ values are significantly impacted by the core repulsion. 

In addition, Table~\ref{tab:C6svtot} contains the values of $\vdvmax$ and $\vmaxone$ for each of these collision partners. As per the discussion, the non-universal species listed here ($^{87}$Rb-He and $^{87}$Rb-Ne) share the characteristic that $\vdvmax > \vmaxone$, indicating that the velocity averaging occurs outside the $C_6$~-~dominated collision region.

\begin{table}[h]
\centering
\begin{tabular}{||c|cc|r|r||}
\hline
& \multicolumn{2}{c|}{($10^{-15}$m$^3$/s)} &\multicolumn{2}{c||}{(m/s)} \\
 Collision Pair &  $\svtotCsix$ & $\svtot_{\rm{LJ}}$ & $\vdvmax$\phantom{ } & $\vmaxone$\phantom{ }\\
\hline
$^{87}$Rb-He & 2.49 & 2.10 &\phantom{ } 1354\phantom{ }& \phantom{ }62\phantom{ }\\
$^{87}$Rb-Ne & 2.01 & 1.72 & \phantom{ }603\phantom{ } &  \phantom{ }281\phantom{ }\\
\hline
$^{87}$Rb-Ar & 2.81 & 2.82 &\phantom{ } 428\phantom{ } & \phantom{ }1638\phantom{ }\\
$^{87}$Rb-Kr & 2.63 & 2.64 &\phantom{ } 296\phantom{ } & \phantom{ }2785\phantom{ }\\
$^{87}$Rb-Xe & 2.75 & 2.76 &\phantom{ } 236\phantom{ } & \phantom{ }4243\phantom{ }\\
\hline
\end{tabular}
\caption{\justifying{Comparison of $\svtot$ values given an interaction potential with and without its short range repulsive barrier for a variety of collision partners.  The $\svtot_{\rm{LJ}}$ are from FQMS computations using a Lennard-Jones potential, while the $\svtotCsix$ are the JB approximation for a purely $C_6$ potential, Eq.~\ref{eq:svtot6JB}.  Note the close agreement between these two values for the heavier collision partners, $^{87}$Rb-Ar, $^{87}$Rb-Kr, and $^{87}$Rb-Xe. The final two columns indicate the values of $\vdvmax$ and $\vmaxone$ for each species. As expected, the universal species share the characteristic that $\vdvmax < \vmaxone$. For all of the entries the ambient temperature was chosen as 294 K for the MB averaging, and the $C_6$ and $D_e$ parameters are taken from \cite{Klos:2023}.}}
\label{tab:C6svtot}
\end{table}



\section{Universality Revisited} \label{TNG}

In the previous sections, the universality hypothesis was illustrated using the LJ potential.  In this section, we examine the validity of the universality hypothesis using more realistic models of the interaction potentials taken from Medvedev \textit{et al} \cite{Medvedev2018} and K{\l}os and Tiesinga \cite{Klos:2023}.  For this purpose, we develop a more accurate semi-classical model (SC) including the other long-range van der Waals interactions, $-C_{8}/r^8$ and $-C_{10}/r^{10}$, that were omitted in prior work and in the preceding analysis.

\subsection{Including $C_8$ and $C_{10}$ long range terms.}\label{sec:OLRT}

The other long-range van der Waals interactions, $-C_{8}/r^8$ and $-C_{10}/r^{10}$, that were omitted in prior work and in the preceding analysis add to the collision phase shifts, $\eta_L(k)$, and, increase the $\sigma_{\rm{tot}}(v)$ and $\svtot$ values.  The phase shift for a purely $C_6$, $C_8$, and $C_{10}$ potential can be written using the Jeffreys-Born (JB) approximation as
\be
\eta_L(k) = \frac{3 \pi}{16} \left(\frac{\mu C_6}{\hbar^2}\right) \frac{k^4}{\ell^5} +\frac{5 \pi}{32} \left(\frac{\mu C_8}{\hbar^2}\right) \frac{k^6}{\ell^7} + \frac{35 \pi}{256} \left(\frac{\mu C_{10}}{\hbar^2}\right) \frac{k^8}{\ell^9}.
\label{eq:etaL6810}
\ee
(Recall, $\ell = L + 1/2$.)
This semi-classical (SC) model contains no core potential in the elastic phase shifts, and is an extension of the purely $C_6$ model, Eq.~\ref{eq:JBetaL6}.

\begin{figure}
\includegraphics[width=0.48\textwidth]{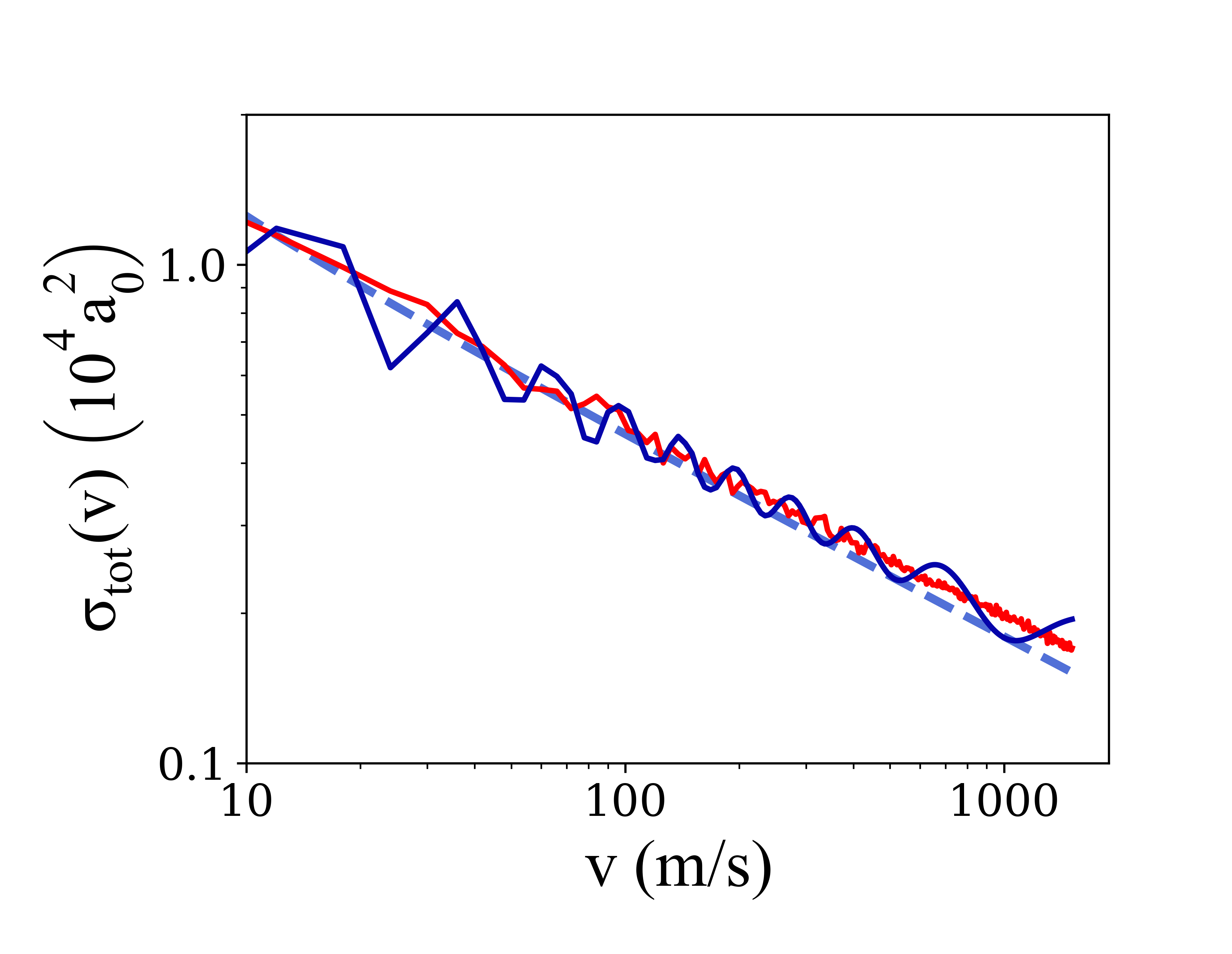} 
\caption{\justifying{Plot of $\sigma_{\rm{tot}}(v)$ versus $v$ for FQMS computations based on the K{\l}os and Tiesinga \cite{Klos:2023} model interactions (blue solid trace), the purely $C_6$ model in Eq.~\ref{eq:svtot6JB} (blue dashed trace), and the SC model of Eq.~\ref{eq:etaL6810} (red trace) for $^{87}$Rb-Ar.  The FQMS and SC predictions rise slightly above the pure $C_6$ predictions owing to the addition of the long-range $C_8$ and $C_{10}$ terms in these models.  This leads to the $\svtot$ value being systematically higher for a potential containing these terms than standard the LJ model.}}
    \label{fig:RbArXeSCcompare}
\end{figure}

To illustrate the effect of these omitted terms, Fig.~\ref{fig:RbArXeSCcompare} shows the $^{87}$Rb-Ar total collision cross-section predictions, $\sigma_{\rm{tot}}(v)$, versus relative collision speed, $v$, based on the SC model (red solid trace) and the purely $C_6$ model (blue dashed trace) compared with those obtained from FQMS calculations using the K{\l}os and Tiesinga inter-species potentials \cite{Klos:2023} (blue solid trace).  

The FQMS and SC predictions rise slightly above the pure $C_6$ predictions owing to the addition of the long-range $C_8$ and $C_{10}$ terms in these models. Notably, the SC model does not display any significant glory undulations, as the core portion of the potential responsible for these is absent. The undulatory FQMS $\sigma_{\rm{tot}}(v)$ values are centered around the SC values, underscoring that the role of the core repulsion is mainly confined to creating the glory oscillations over the range of $v < \vmaxone$ depicted.


\subsection{Limitations of the Universality approximation}\label{sec:Limitations}

The minimization of the short range interaction potential influence on the collision rate coefficient is the result of velocity averaging for $\vdvmax < \vmaxone$ which renders $\svtot$ insensitive to the details of the glory undulations.  The result is a value of $\svtot$ which is well approximated by the SC estimate despite the SC model providing no account of the short range whatsoever.  In the following, we examine the accuracy of the approximation, $\svtot \simeq \svtotSC$, for different sensor atoms and a variety of background gases.

We note at the outset of this discussion that the JB approximation upon which the SC phase shifts (and predictions) are based is not expected to be good at low velocities or small values of the angular momentum, $L$.  However, the generic trend of $\sigma_{\rm{tot}}(v)$ seems to be captured by the SC model, and so the discrepancies of the $\svtot$ and $\svtotSC$ predictions will serve here as a heuristic indicator of universality.  A full quantitative study of the actual insensitivity of $\svtot$ to changes in the interaction potential (e.g.~the values of $C_6, \mu, D_e$) is the subject of future work \cite{Xuyang2024}.

Residual sensitivity to the interaction potential at short range and the resulting deviations of $\svtot$ from $\svtotSC$ are expected even when the velocity average is restricted to the universal regime, $\vdvmax < \vmaxone$. These discrepancies are more significant when the glory undulations have a large amplitude and the velocity averaging does not encompass {an integral number of undulations} in the computation of $\svtot$. For lower mass sensor atoms, the number of partial waves included in the computation of $\sigma_{\rm{tot}}(v)$ is reduced. As a result, the effects of glory scattering -- which produce constructive or destructive interference centered around a partial wave, $L_g$ (see Appendix A) -- have a relatively larger impact on the cross-section. This results in glory undulations with a larger amplitude. In addition, Bernstein  \cite{10.1063/1.1733383,10.1063/1.1733558} and Child \cite{child1996molecular} assert that the number of glory undulations that appear in $\sigma_{\rm{tot}}(v)$ equals the number of rotation-less bound states that the inter-species potential can support. Lowering the reduced mass of the collision pair for a fixed potential depth lowers the number of these states, consequently reducing the number of glory oscillations in the $\sigma_{\rm{tot}}(v)$ versus $v$ spectrum. Both of these factors contribute to rendering the velocity averaging less effective in removing the influence of the glory scattering on the value of $\svtot$.

\begin{center}
\begin{figure*}
   \centering
    \begin{subfigure}[t]{0.48\textwidth}
        \centering
        \includegraphics[width=\linewidth]{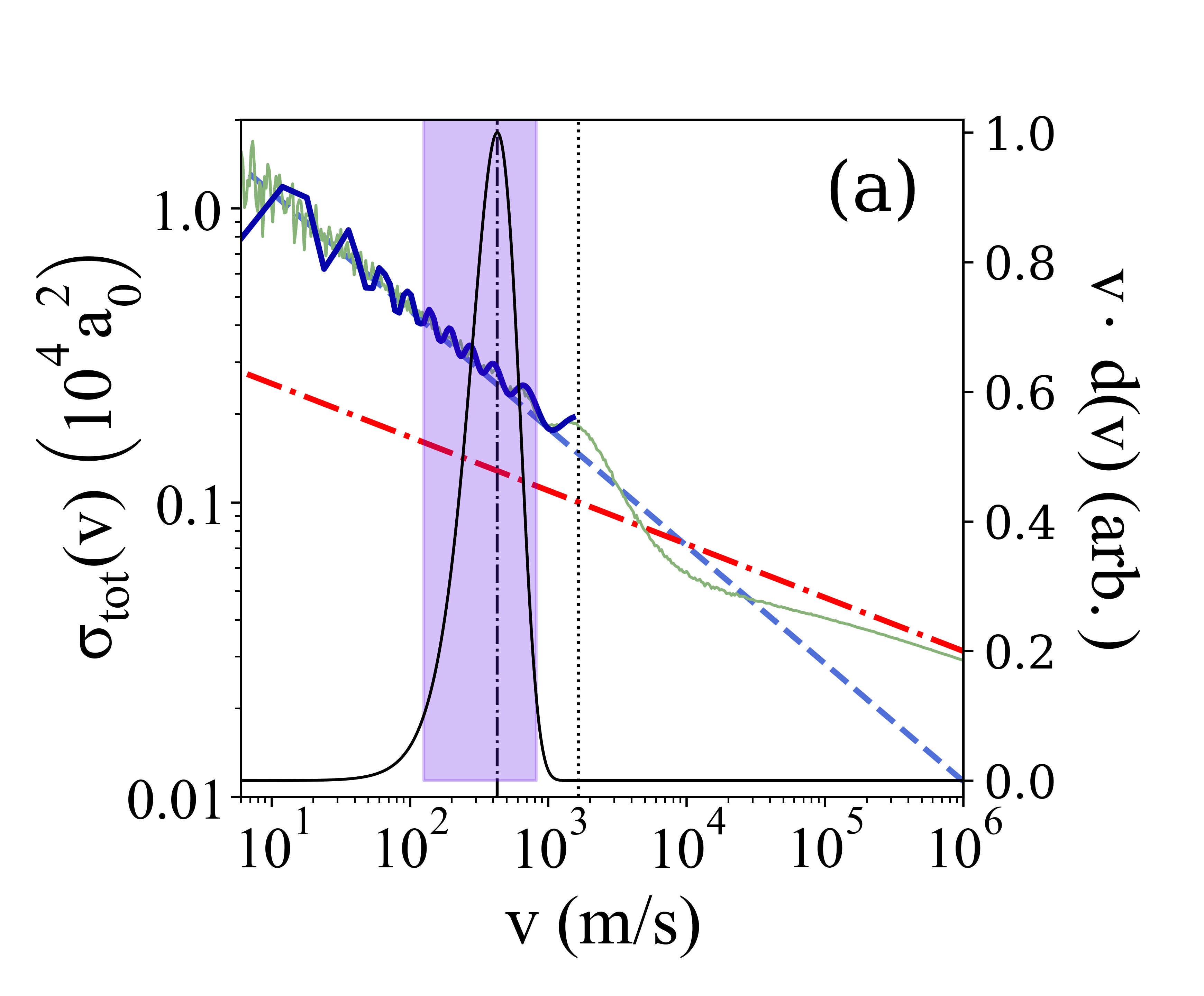} 
         \label{fig:RbArsig}
   \end{subfigure}
   \centering
       \begin{subfigure}[t]{0.48\textwidth}
       \centering
        \includegraphics[width=\linewidth]{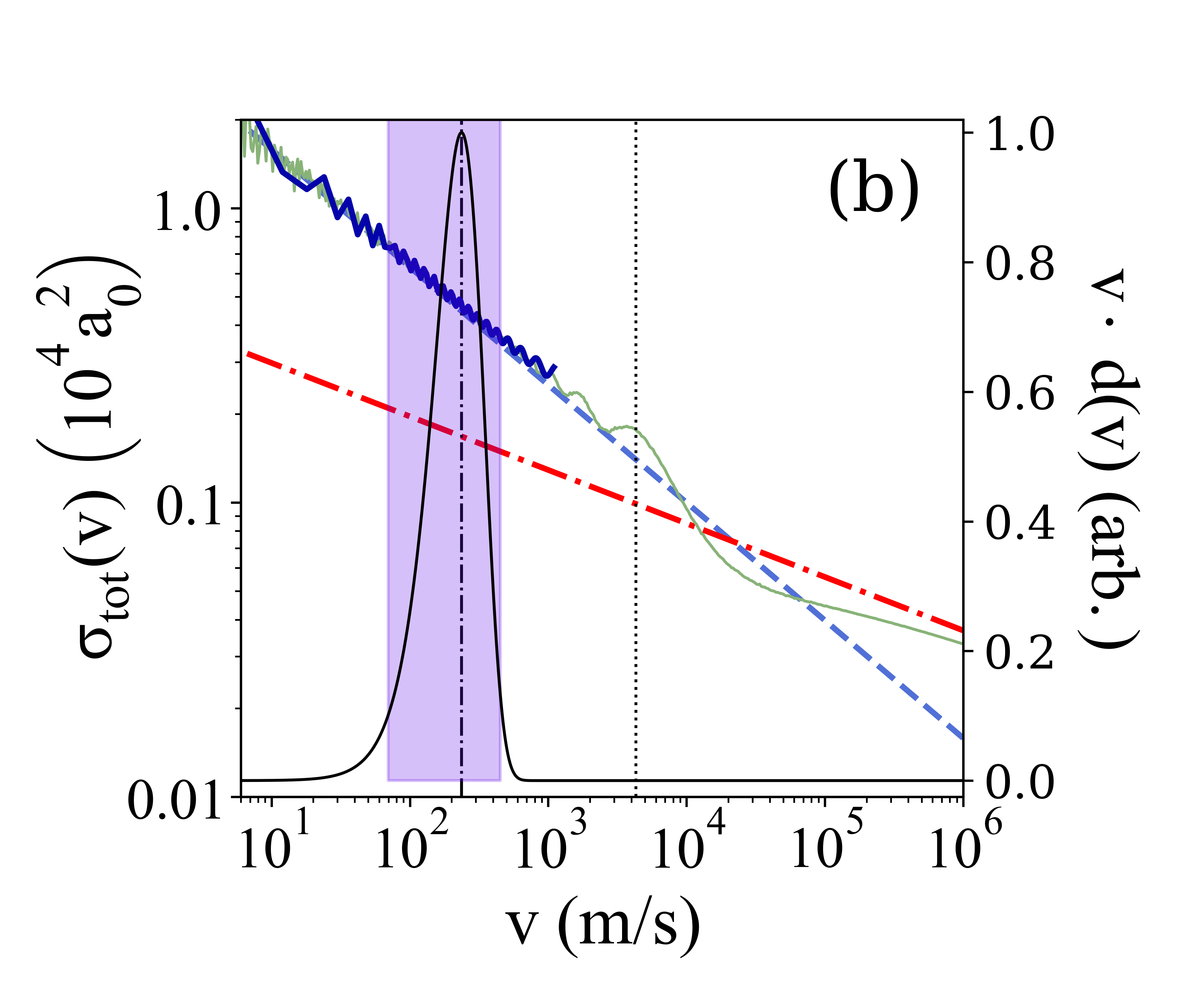} 
        \label{fig:RbHesig}
   \end{subfigure}

   \centering
    \begin{subfigure}[t]{0.48\textwidth}
        \centering
        \includegraphics[width=\linewidth]{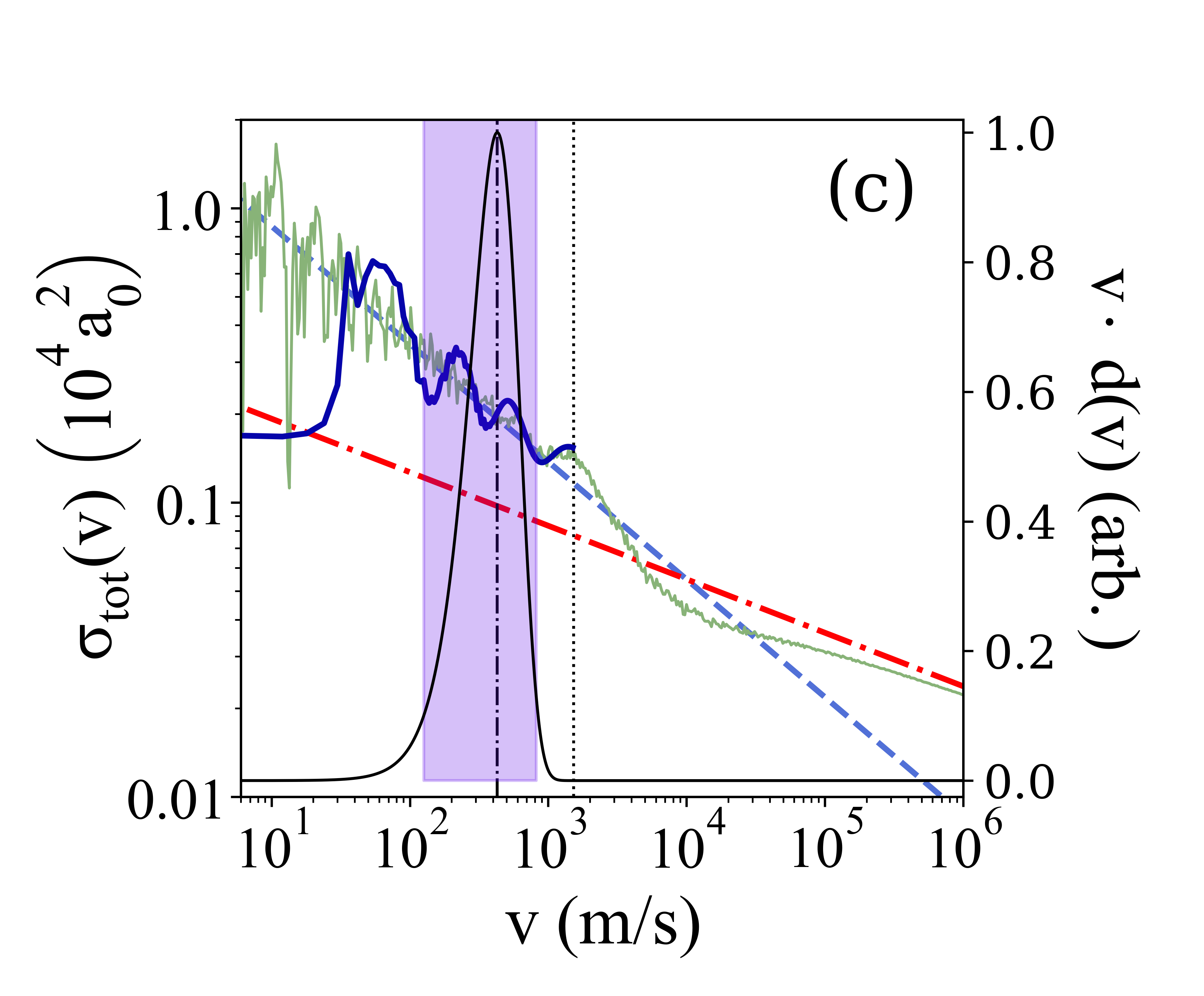} 
         \label{fig:RbArsig}
   \end{subfigure}
   \centering
       \begin{subfigure}[t]{0.48\textwidth}
       \centering
        \includegraphics[width=\linewidth]{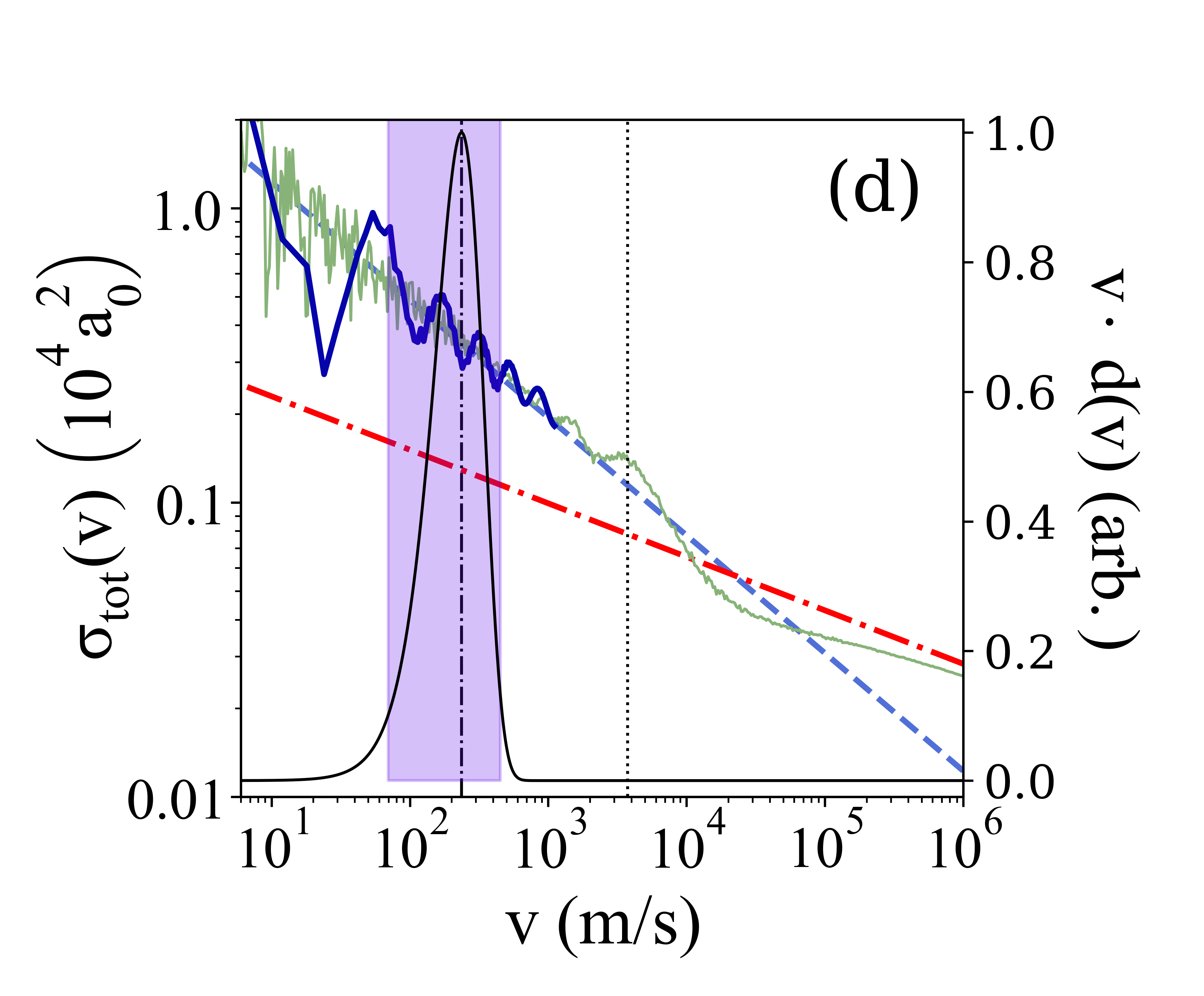} 
        \label{fig:RbHesig}
   \end{subfigure}
    \caption{\justifying{Plots of the $\sigma_{\rm{tot}}$ versus $v$ for (a) $^{87}$Rb-Ar, (b) $^{87}$Rb-Xe, (c) $^{7}$Li-Ar, and (d) $^{7}$Li-Xe for different interaction potential models.  The cross section values are shown for FQMS computations using the KT potential \cite{Klos:2023} (thick blue traces), for a pure long-range $C_6$ potential, Eq.~\ref{eq:stotJBvall} (blue dashed traces),
    and for a pure short-range $C_{12}$ potential predictions, Eq.~\ref{eq:stot12JB} (red dot-dashed traces). The thin green traces show $\sigma_{\rm{tot}}(v)$ computed using an augmented Lennard-Jones (a-LJ) potential which incorporates the $C_8$ and $C_{10}$ long-range van der Waals terms. The $C_n$ values values were taken from K{\l}oss and Tiesinga (KT) \cite{Klos:2023} and the $C_{12}$ value was fixed to insure that the potential depth matched the KT potential value of $D_e$.  Note how both the FQMS and modified LJ results rise above the purely $C_6$ predictions owing to the contributions of the $C_8$ and $C_{10}$ terms. 
The black dotted vertical line on each plot indicates $\vmaxone$, above which the cross-sections transition from long-range to short-range character. Also shown are the normalized velocity weighted MB velocity distributions, $v\cdot d(v)$, (black solid traces) with their peak speeds indicated by the vertical dash-dotted lines for the room-temperature ($T=294$~K) background gas collision partners. The shaded regions indicate the range over which the normalized $v\cdot d(v) \ge 0.1$.}}
    \label{fig:RbArXeLiArXecompare}
\end{figure*}
\end{center}

These effects are illustrated in Fig.~\ref{fig:RbArXeLiArXecompare}. Figs.~\ref{fig:RbArXeLiArXecompare}(a) and (b) depict the $^{87}$Rb-Ar and $^{87}$Rb-Xe $\sigma_{\rm{tot}}(v)$ vs $v$ values, respectively. These are contrasted by Fig.~\ref{fig:RbArXeLiArXecompare}(c) and (d) which show the $^{7}$Li-Ar and $^{7}$Li-Xe computations, respectively. On each figure the FQMS computations (blue solid traces) based on the KT potentials \cite{Klos:2023}, the purely $C_6$ predictions (blue dashed traces) and purely $C_{12}$ predictions (red dash-dot trace) are shown. The green traces are the result of an augmented Lennard-Jones (a-LJ) potential model which includes the long-range $C_8$ and $C_{10}$ van der Waals terms (the $C_n$ values were taken from \cite{Klos:2023}).
\bea 
V_{\rm{a\text{-}LJ}}(r) &=& \frac{C_{12}}{r^{12}} - \frac{C_{10}}{r^{10}} - \frac{C_8}{r^8} - \frac{C_6}{r^6}.
\label{eq:aLJ}
\eea
The $C_{12}$ repulsive core coefficients were chosen such that the potential depths, $D_e$, equaled the values reported by by K{\l}os and Tiesinga \cite{Klos:2023}. The cross-sections for this augmented LJ model were computed using the JB elastic phase shifts (Eqs.~\ref{eq:JBeta12} and \ref{eq:etaL6810}). 

The FQMS computations appear to follow the a-LJ traces at large velocities where the JB approximation is expected to be good, and both rise above the purely $C_6$ prediction because they include the $C_8$ and $C_{10}$ long-range terms.  This underscores the message of section \ref{sec:OLRT} that the presence of the other long-range terms ($C_8$ and $C_{10}$) systematically increases the $\svtot$ value above the $\svtotCsix$ prediction.  In addition, it is clear that the $^{7}$Li-X (X = Ar or Xe) demonstrate fewer, larger amplitude glory undulations compared to their $^{87}$Rb-X (X = Ar or Xe) counterparts. Thus, the corresponding $\svtot$ values for $^{7}$Li-X cases are expected to exhibit a larger deviation from $\svtotSC$, the predictions of the undulation-free SC model.  The comparisons shown in Table~\ref{tab:svtotLiRb} are consistent with this expectation.


To assess the validity of the universality approximation for $^{87}$Rb-X collisions, FQMS computations of $\svtot$ were carried out using the interaction potential models of Medvedev \textit{et al} \cite{Medvedev2018}, furnishing $\svtot_{\rm{Med}}$, and of K{\l}os and Tiesinga \cite{Klos:2023}, providing $\svtot_{\rm{KT}}$.  Similarly, semi-classical (SC) values,
$\svtotSC$, were computed along with the purely $C_6$ predictions, $\svtotCsix$, Eq.~\ref{eq:svtot6JB}. For all of these computations the MB averaging was carried out at an ambient temperature of 294~K. The results are presented in Table~\ref{tab:svtotall}.
\begin{table}[t]
\centering
\begin{tabular}{||c|c|c|cc||}
\hline
& \multicolumn{4}{c||}{($10^{-15}$m$^3$/s)} \\
Collision Pair &  $\svtotCsix$ & $\svtot_{\rm{SC}}$ & $\svtot_{\rm{Med}}$ & $\svtot_{\rm{KT}}$  \\
\hline
$^{87}$Rb-He & 2.49 & 3.08 & 2.44 & 2.37 \\
$^{87}$Rb-Ne & 2.01 & 2.28 & 1.98 & 2.04 \\
\hline
$^{87}$Rb-Ar & 2.81 & 3.00 & 3.04 & 3.02 \\
$^{87}$Rb-Kr & 2.63 & 2.77 &2.79 & 2.78 \\
$^{87}$Rb-Xe & 2.75 & 2.88 & 2.88 & 2.87 \\
\hline
$^{87}$Rb-Rb  & 6.38 & 6.58 & --- & --- \\
\hline
\end{tabular}
\caption{\justifying{A comparison of the computed $\svtot$ values for $^{87}$Rb-X (X = He, Ne, Ar, Kr, and Xe) at T = 294 K.  The predictions based on FQMS calculations using the potentials from \cite{Medvedev2018} are labeled $\svtotMED$ and those obtained using the potentials from \cite{Klos:2023} are labeled $\svtotKT$. The semi-classical predictions, $\svtotSC$, are remarkably close to the FQMS calculations for the heavy collision partners although this SC calculation is based on the Jeffreys-Born approximation (Eq.~\ref{eq:svtot6JB}) for the long-range induced scattering phase shifts and contains no information about the core repulsion of the potential.  The purely $C_6$ predictions, $\svtotCsix$, are distinct from the $\svtotSC$ predictions because the former does not include the phase shifts from the other long-range potential terms (i.e.~the $C_8$ and $C_{10}$ terms). The final row shows the corresponding $^{87}$Rb-Rb collision $C_6$ and SC values, based on the $C_n$ from \cite{Mitroy2015}.}}
\label{tab:svtotall}
\end{table}

For the universal pairs, $^{87}$Rb-Ar, $^{87}$Rb-Kr, and $^{87}$Rb-Xe, the results are noteworthy: The $\svtotMED$, $\svtotKT$, and $\svtotSC$ values are equivalent at the level of $<$ 1\%.  Because the KT and Med potentials are distinct at short range and because the SC calculation is devoid of any short range effects whatsoever, one concludes that the $\svtot$ values for the universal pairs are independent of the interaction potential at short range due to MB averaging.  In addition, the $\svtotCsix$ values are systematically lower than the corresponding $\svtotMED$, $\svtotKT$, and $\svtotSC$ values.  Clearly, the $C_8$ and $C_{10}$ values contribute a small, but non-negligible amount to the total collision cross-section coefficients (7\% for $^{87}$Rb-Ar, 5\% for $^{87}$Rb-Kr, 4\% for $^{87}$Rb-Xe, and 3\% for $^{87}$Rb-Rb). 
By contrast, the non-universal pairs, $^{87}$Rb-He and $^{87}$Rb-Ne, display large discrepancies between the SC predictions, $\svtotSC$, and the FQMS physical model values, $\svtotMED$ and $\svtotKT$.  Notably, the FQMS values also differ from each other due to the differences in their short range interaction potential while they have the same shape at long-range.  This is a manifestation of the breakdown of universality for light collision partners as observed with direct measurements in 2023 \cite{Shen_2023}.

To investigate the quality of the SC approximation further, the total collision rate coefficients for $^{7}$Li and $^{87}$Rb sensor atoms colliding with both atomic and molecular species (H$_2$, He, Ne, N$_2$, Ar, Kr, and Xe) reported by K{\l}os and Tiesinga (KT) in Ref~\cite{Klos:2023}, $K(T) = K_0 + K_1(T-300 \rm{K})$, are compared to values obtained using the SC approximation, $\svtotSC$, using the KT model $C_n$ values. The results are shown in Table~\ref{tab:svtotLiRb}. The agreement is remarkably good for the (universal) heavier collision partners, N$_2$, Ar, Kr, and Xe.
The SC and $K(294 \rm{K})$ values for $^{7}$Li-X (X = N$_2$ and Ar) agree within the uncertainty range and the $^{7}$Li-(Kr and Xe) values agree within $<$ 2\%. Similarly, the $^{87}$Rb-X (X = N$_2$, Kr, Xe) values agree within 0.5\%, while the $^{87}$Rb-Ar results agree within 1.1\%.  Thus, the ansatz that the $\svtot$ values are independent of the details of the core potentials appears to be supported for both $^{7}$Li and $^{87}$Rb sensor atoms. For the lighter collision partners examined, H$_2$, He, and Ne, the center of their $v \cdot d(v)$ distribution, $v_{vdv}^{\rm{max}}$, lies above the estimated long range dominated velocity limit, $\vmaxone$.  The result is that $\svtot$ is influenced by the core repulsion and the SC prediction is different from the FMQS value by up to 50\%.

\begin{table}[t]
\centering
\begin{tabular}{||c|cc|c||}
\hline
 & \multicolumn{2}{c|}{($10^{-15}$m$^3$/s)}&\  $\svtot_{\rm{SC}}$\ \\
 Collision Pair &  $K(294 \rm{K})$ \cite{Klos:2023} & $\svtot_{\rm{SC}}$ &\  $\overbar{K(294 \rm{K})}$\ \\
\hline
$^{7}$Li-H$_2$ &    3.16(6) & 4.46 &  1.41(3) \\
$^{7}$Li-He &    1.65(4) & 2.31 &  1.40(3) \\
$^{7}$Li-Ne &    \phantom{1}1.56(14) & 1.71 &  \phantom{2}1.10(10) \\
\hline
$^{7}$Li-N$_2$ & 2.63(2) & 2.63 &  \phantom{2}0.999(8) \\
$^{7}$Li-Ar &    \phantom{2}2.329(6) & 2.32 &  \phantom{2}0.996(3) \\
$^{7}$Li-Kr &    \phantom{2}2.145(4) & 2.11 & \phantom{2}0.986(2) \\
$^{7}$Li-Xe &    2.24(2) & 2.21 &  \phantom{2}0.988(9) \\
\hline 
\hline
$^{87}$Rb-H$_2$ & \phantom{2}3.88(10) & 5.89 &  1.52(4) \\
$^{87}$Rb-He &    2.37(3) & 3.08 &  1.30(2) \\
$^{87}$Rb-Ne &    2.0(2)\phantom{2} & 2.28 &  \phantom{2}1.14(12) \\
\hline
$^{87}$Rb-N$_2$ & 3.45(6) & 3.45 & \phantom{2} \phantom{2}1.001(17)\phantom{2} \\
$^{87}$Rb-Ar &    \phantom{2}3.031(7) & 3.00 &  \phantom{2}0.989(2) \\
$^{87}$Rb-Kr &    2.78(1) & 2.77 &  \phantom{2}0.996(4) \\
$^{87}$Rb-Xe &    2.87(1) & 2.88 & \phantom{2}1.003(4) \\
\hline
\end{tabular}
\caption{\justifying{A comparison of the total collision rate coefficient values, $K(T) = K_0 + K_1(T~-~300~ \rm{K})$, 
\phantom{2} from Ref.~\cite{Klos:2023} 
for $^{87}$Rb-X and $^{7}$Li-X (X~=~H$_2$, He, Ne, N$_2$, Ar, Kr, and Xe) at T = 294K
to the semi-classical estimates computed here, $\svtot_{\rm{SC}}$. The final column gives the ratio of the two. The uncertainties quoted in the ratios are derived from the uncertainties listed in column 2.  Remarkably, the SC predictions and the $K(294 \rm{K})$ values are in very good agreement for $^{87}$Rb~-~X  and $^{7}$Li~-~X for the heavy collision partners, X = N$_2$, Ar, Kr, and Xe.}}
\label{tab:svtotLiRb}
\end{table}

A notable feature of the results in Table~\ref{tab:svtotLiRb} is that they suggest $\svtot$ may be insensitive to the anisotropy of the interactions.  In the case of diatomic molecule-atom collisions, such as $^{87}$Rb-N$_2$, the anisotropic interactions are appreciable compared to the isotropic interaction.  This results in both elastic and inelastic collisions.  The values for the total collision rate coefficient (including both elastic and inelastic collisions) reported by K{\l}os and Tiesinga \cite{Klos:2023} and shown in Table~\ref{tab:svtotLiRb} were obtained using a full multi-channel scattering calculation and the complete interaction potentials including both the isotropic and anisotropic parts.  Remarkably, the values for $\svtot$ obtained using the semi-classical approximation that neglects both the anisotropy and the short range interaction potential are in near perfect agreement with those obtained using the complete interaction potential for the diatomic molecule-atom collisions $^{87}$Rb-N$_2$ and $^{7}$Li-N$_2$.  This agreement could be accidental; however, a numerical study quantifying the sensitivity of $\svtot$ to changes in the interaction potential has recently been submitted \cite{Xuyang2024}, and preliminary results extending that work to atom-molecule collisions show evidence that $\svtot$ is, indeed, invariant to changes of the anisotropic part of the interaction potential \cite{Xu}.  A somewhat related experimental result is that no difference in the ratios of $\svtot$ for these two sensor atoms when exposed to ortho-H$_2$ versus para-H$_2$ was observed \cite{Frieling2024AVS}.

\section{Revising $\pqdu$}\label{Refine}

\begin{figure}
   \centering    
   \includegraphics[width=0.5\textwidth]{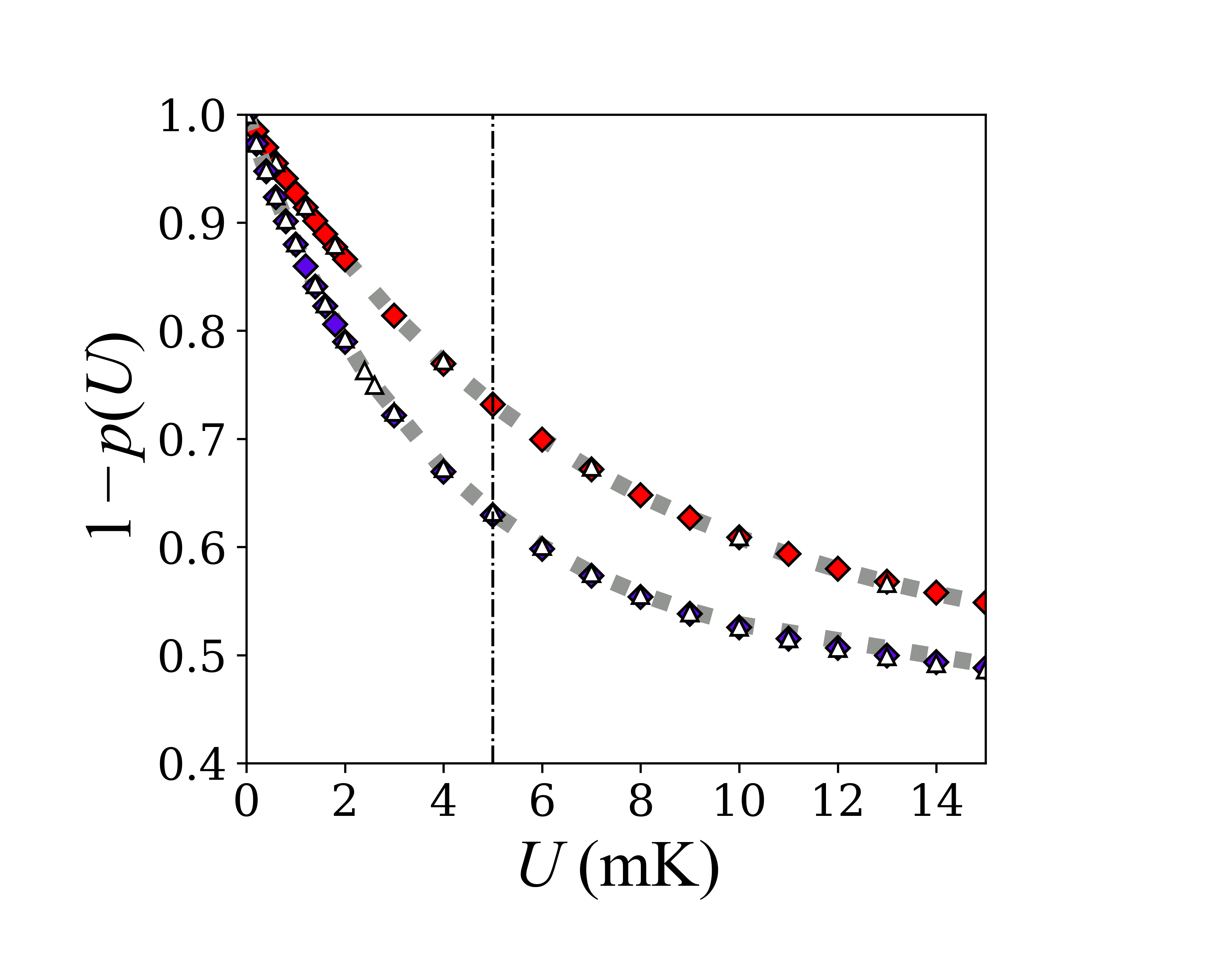} 
    \caption{\justifying{A plot of $\svloss/\svtot = 1 - p(U)$ versus trap depth, $U$ for collisions between $^{87}$Rb and the heavy (universal) partners, Ar and Xe, at T = 294K. The grey dotted traces are the FQMS prediction using the K{\l}os and Tiesinga (KT) potetial \cite{Klos:2023} for $^{87}$Rb-Ar (upper trace) and for $^{87}$Rb-Xe (lower trace). The SC prediction model for $^{87}$Rb-Ar is shown as red diamonds, and the SC prediction for $^{87}$Rb-Xe is shown as the violet diamonds. Overlaid on these are the corresponding FQMS predictions for the LJ model (white triangles). Note how the $^{87}$Rb-Ar and $^{87}$Rb-Xe predictions are clearly distinct, but the predictions from using different interaction potentials overlap well over this range of trap depths, indicating that they yield the same $p(U)$.
    The vertical dot-dashed line is the maximum trap depth attained thus far in experiments.}}
    \label{fig:scaledsvlossvsU}
\end{figure}

The trap loss rate coefficient, $\svloss$ (Eq.~\ref{eq:svlosspU}), was previously studied using FQMS calculations with a LJ interaction potential and was found to be insensitive to the shape of the interaction potential at short range.  The shape of $\svloss/\svtot$ was then fit to a polynomial whose coefficients are provided in Table~\ref{tab:betajs}.  Because the LJ model used lacks the $C_8$ or $C_{10}$ long-range terms, the validity of this prior analysis is suspect.  To investigate this further, FQMS computations of $\svloss$ were carried out using the more realistic K{\l}os and Tiesinga (KT) potential model \cite{Klos:2023}, the LJ model, and the SC model for the universal collision pairs $^{87}$Rb-Ar and $^{87}$Rb-Xe. 
Fig.~\ref{fig:scaledsvlossvsU} shows $1-p(U)$ versus $U$ for these universal collision pairs. While the shapes of the curves differ for the different collision pairs, $^{87}$Rb-Ar and $^{87}$Rb-Xe, the shapes for each pair are independent of the collision model used, for shallow traps. That is, for $^{87}$Rb-Ar the values for the SC model (red diamonds) and the LJ model (white triangles) overlap the realistic KT model (grey dotted trace). Similarly, the SC model (violet diamonds) and LJ model (white triangles) values overlap the KT model (grey dotted trace) for $^{87}$Rb-Xe in Fig.~\ref{fig:scaledsvlossvsU}. 

\begin{figure}[h!]
   \centering
    \begin{subfigure}[t]{0.48\textwidth}
        \centering
        \includegraphics[width=\linewidth]{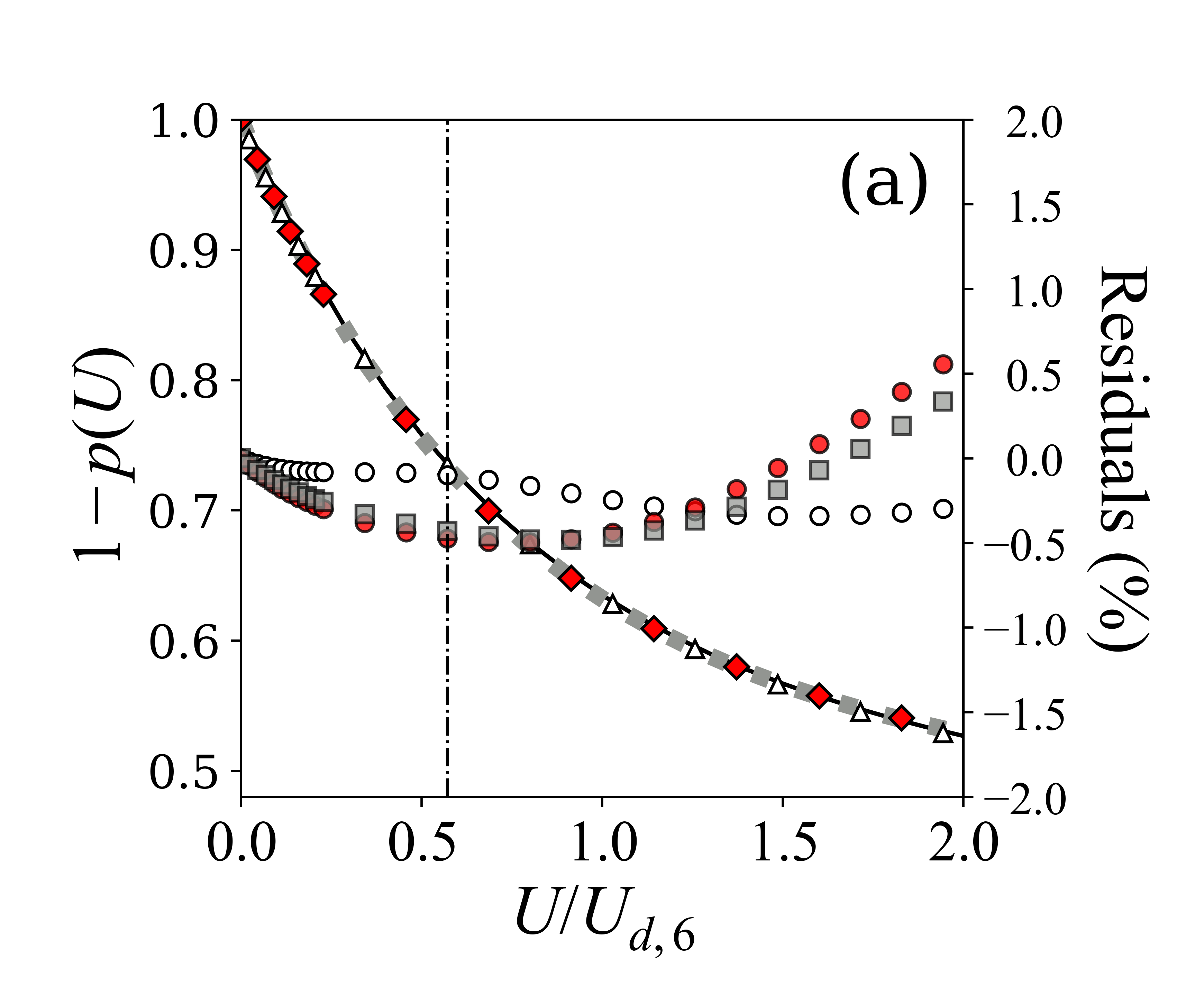} 
         \label{fig:RbArsig}
   \end{subfigure}
   \centering
       \begin{subfigure}[t]{0.48\textwidth}
       \centering
        \includegraphics[width=\linewidth]{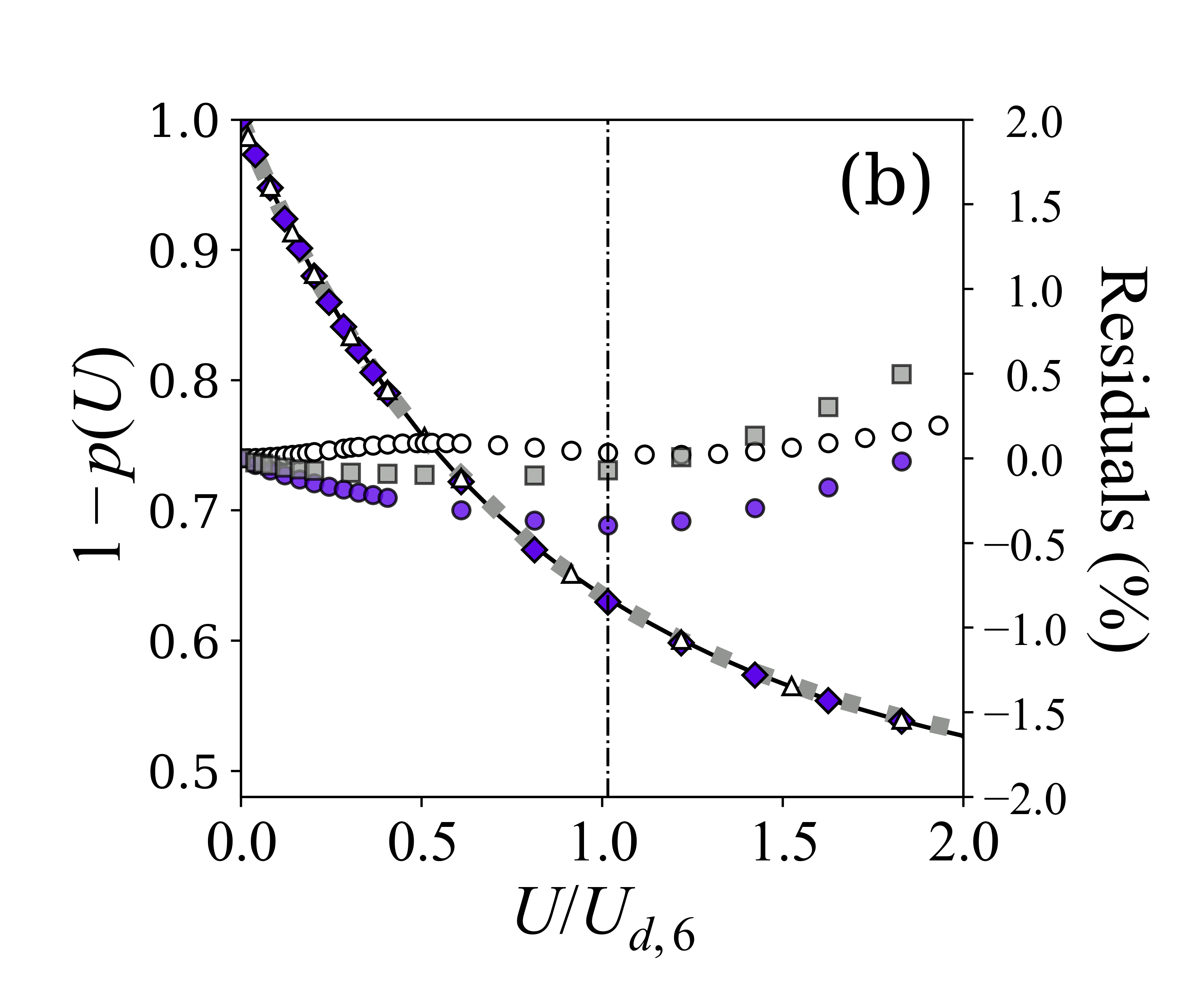} 
        \label{fig:RbHesig}
   \end{subfigure}
   \vspace{-5ex}
\caption{\justifying{
Plots of $\svloss/\svtot = 1 - p(U)$ versus the scaled trap depth, $U/\Udsix$, using different potential models for collisions between $^{87}$Rb and Ar and between $^{87}$Rb and Xe (universal species) at T = 294 K.  The overlap between the predictions is remarkable: (a) shows $^{87}$Rb-Ar FQMS values based on the K{\l}os and Tiesinga (KT) model (dotted gray trace), the SC model (red diamonds),  and the LJ model (white triangles). (b) shows the $^{87}$Rb-Xe KT model (dotted grey trace), SC model (violet diamonds), and LJ model (white triangles). The universal predictions from \cite{Booth2019, Shen_2020} are shown in both figures as the solid black trace. The percent residuals between the different models and the universal expression, Eq.~\ref{eq:pqdu6}, are plotted as individual points on each figure: KT model residuals as grey  squares, SC model residuals as red circles for $^{87}$Rb-Ar and as violet circles for $^{87}$Rb-Xe, while the white circles are the LJ residuals on both plots.  The vertical dot-dashed lines are the maximum achievable $U/\Udsix$ attained thus far in experiments.}}
    \label{fig:RbArXeNISTcompare}
\end{figure}

Thus, one concludes that $p(U)$ depends on the collision pair but does not depend on the collision model employed in the FQMS.  Of key importance is that this comparison shows that $p(U)$ does not depend on the presence of the $C_8$ and $C_{10}$ terms in the potential since they are completely absent for the LJ model, nor does it depend on the details of the core of the potential, as the SC model does not have any repulsive core.  By contrast, the value of $\svtot$ definitely depends on these $C_8$ and $C_{10}$ terms for these universal collision pairs (as discussed in section \ref{sec:OLRT}).
That the three models, sharing only the same long range $C_6$ behavior, predict the same curve yet have distinct total cross sections leads to the prediction that the universal function, $p(U)$, previously defined purely on the basis of the LJ model computations (Eq.~\ref{eq:pqdu}) should persist for these species ($^{87}$Rb-Ar, $^{87}$Rb-Kr, and $^{87}$Rb-Xe), but requires a key modification,
\be
p(U) = p_{\rm{QDU,6}} = \sum_j \beta_j \left(\frac{U}{U_{d,6}}\right)^j.
\label{eq:pqdu6}
\ee
Here,
\be
U_{d,6} = \frac{4 \pi \hbar^2}{m_t \svtotCsix/v_p}
\label{eq:Ud6}
\ee 
and $\svtotCsix$ is the total collision cross-section coefficient based solely on the $C_6$ interaction, Eq.~\ref{eq:svtot6JB}. 
This formulation updates the energy scaling for the normalized trap loss rate to one that only depends on the long-range $C_6$ behavior instead of the total collision rate coefficient.  To illustrate this, Figure~\ref{fig:RbArXeNISTcompare} shows the same data sets but now $1 - p(U)$ is plotted versus the scaled trap depth, $U/U_{d,6}$. The KT model FQMS computations, the LJ values, and the SC predictions converge to the same universal shape previously reported in \cite{Booth2019, Shen_2020}.

\begin{figure}[h!]
   \centering
    \begin{subfigure}[t]{0.48\textwidth}
        \centering
        \includegraphics[width=\linewidth]{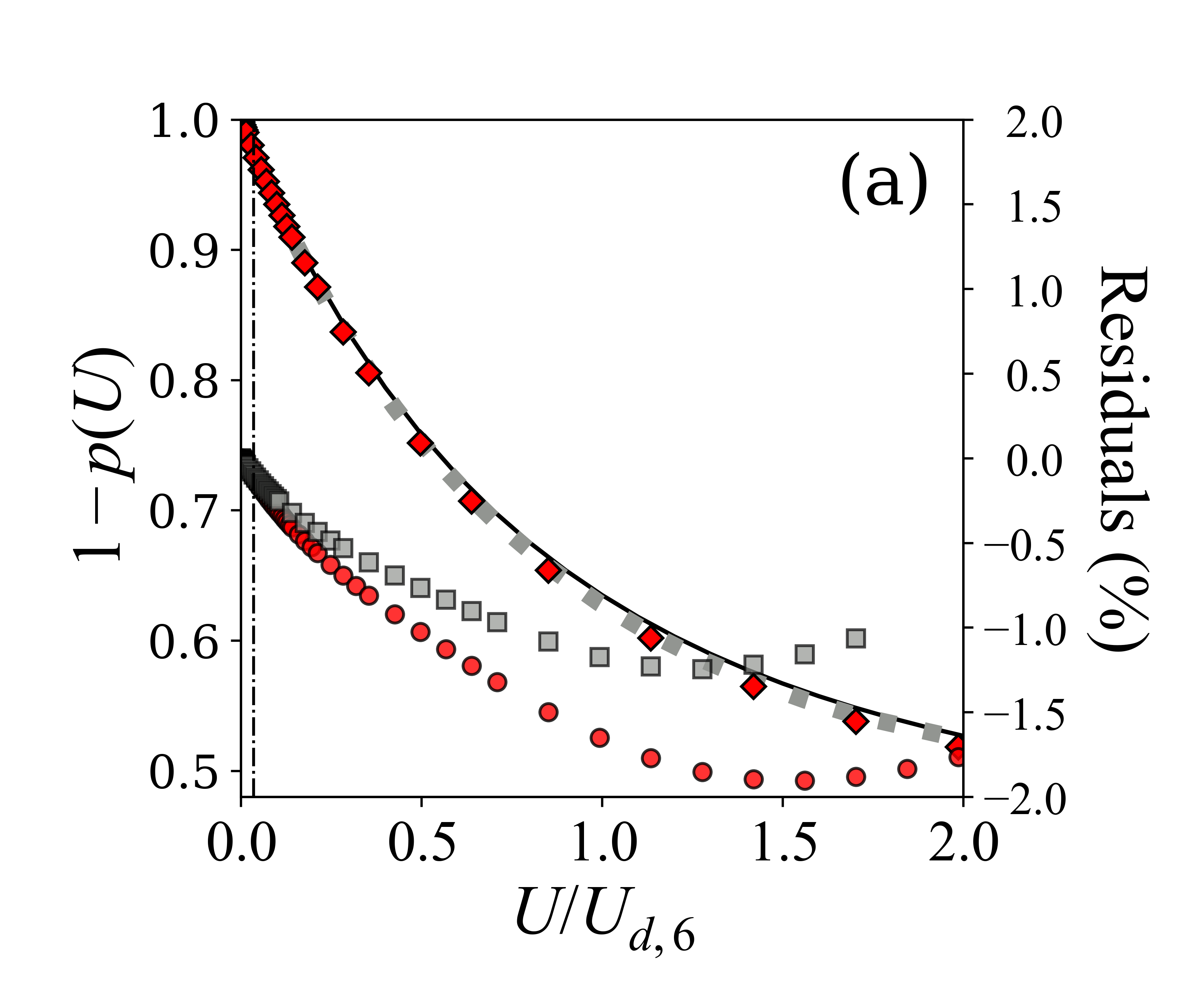} 
   \end{subfigure}
   \centering
       \begin{subfigure}[t]{0.48\textwidth}
       \centering
        \includegraphics[width=\linewidth]{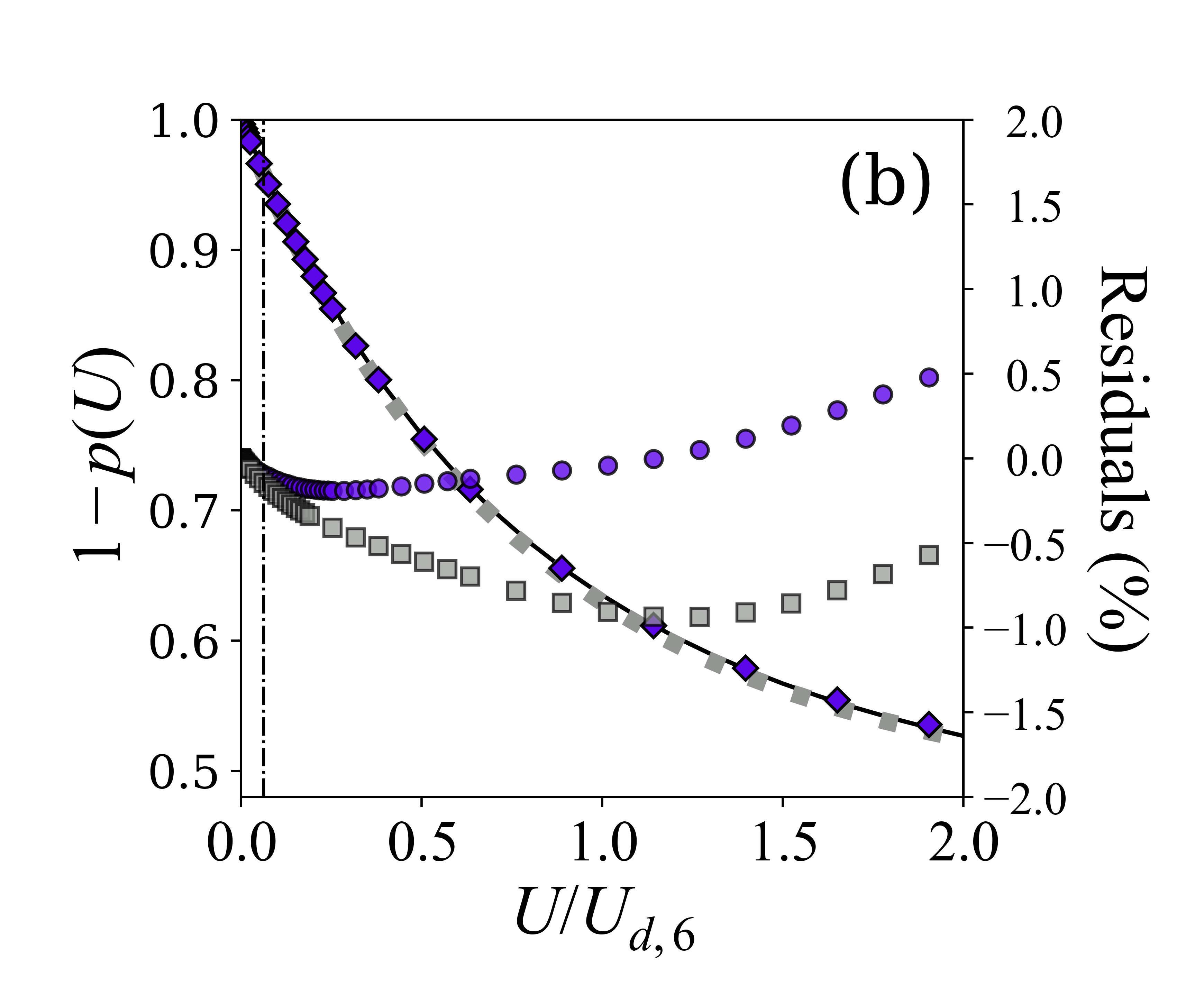} 
   \end{subfigure}
    \caption{\justifying{Plots of $\svloss/\svtot = 1 - p(U)$ versus $U/U_{d,6}$ for (a) $^{7}$Li-Ar and (b) $^7$Li-Xe.  In (a) the dotted grey trace is the FQMS prediction based on the KT model and the red diamonds are the SC model values. The percent residuals between the KT model and universal predictions (grey squares) and between the SC model and universal predictions (red circles). Plot (b) shows the $^{7}$Li-Xe KT FQMS values (grey dotted trace) and the SC model predictions (violet diamonds). In addition the residuals between the KT FQMS values and the universal predictions (grey squares) and between the SC model and the universal model (violet circles) are plotted. The solid black traces are the universal predictions, Eq.~\ref{eq:pqdu6} from \cite{Booth2019, Shen_2020}. These plots underscore that the variation of $1-p(U)$ as a function of $U$ is determined primarily by the long range $C_6/r^6$ portion of the interaction potential for shallow traps. The vertical dot-dashed lines are the maximum achievable $U/\Udsix$ attained thus far in experiments.}}
    \label{fig:LiRbscaledsvlossvsUUd6}
\end{figure}
We find that this updated universal expression for the trap loss rate, Eq.~\ref{eq:pqdu6}, leads to discrepancies with the KT FQMS values and the SC values which are less than 0.5\% for $U/\Udsix \le 1.0$ for $^{87}$Rb-Ar and $^{87}$Rb-Xe collisions. The $\Udsix$ model was also applied to $^{7}$Li-Ar, $^{7}$Li-Kr, and $^{7}$Li-Xe collision pairs (Fig.~\ref{fig:LiRbscaledsvlossvsUUd6}(a) and (b) show the $^{7}$Li-Ar and $^{7}$Li-Xe results). In these cases we observe that the KT FQMS residuals with the universal predictions are systematically negative, up to -1.0\%, while the SC model residuals are also systematically negative, up to -1.7\% over the same $U/\Udsix \le 1.0$ range. The $^{7}$Li-Xe demonstrate the same systematic negative residual between the KT FQMS and universal predictions, while the SC model residuals with the universal function  vary from -0.2\% to 0.5\%.  We limit the range for this consideration out to $U/\Udsix \le 1.0$ because we are using a truncated polynomial approximation to the universal function, and its accuracy is expected to worsen for $U/\Udsix > 1.0$.   This is evident in Fig.~\ref{fig:RbArXeNISTcompare} where the discrepancies grow past $\Udsix$. While the SC and KT FQMS predictions are extremely close for $^{87}$Rb-X collisions, there is a larger discrepancy between them for $^{7}$Li-X.  
The JB approximation is less accurate for lower $L$ collisions, consequently, the SC approximation will be less accurate for $^{7}$Li-X collisions as there are fewer partial waves at play in the collisions.  As can be seen in Fig.~\ref{figA:RbArglory}, $\sigma(v \simeq \vp)$ has significant contributions for partial waves up to $L=35$ for $^{7}$Li-Ar versus up to $L=175$ for $^{87}$Rb-Ar.  The remaining (small amplitude) structure of the FQMS residuals is distinct for each collision pair suggesting that the normalized trap loss rate may not be completely determined by $C_6$ alone; however, understanding the exact origin of these differences is beyond the scope of this work. 


The interpretation of the normalized trap loss rate variation with trap depth following a universal curve is that quantum diffractive collisions, which impart very little momentum and energy to the trapped ensemble and contribute to sensor atoms being retained at small ($<1$~mK) trap depths, primarily probe the long range $-C_6/r^6$ portion of the interaction potential. In the classical sense, these collisions correspond to very large impact parameter collisions and thus only probe the interaction potential at extremely long range.

Because $p(U)$ for shallow traps ($U \le \Udsix$) only depends on $U_{d,6}$, the atom sensor is not self-calibrating as originally conceived. That is, fitting the experimentally measured loss rate ratio, $\Gamma_{\rm{loss}}(U)/\Gamma_{\rm{loss}}(U = 0)$ at a fixed background gas pressure for different trap depths, $U$, to Eq.~\ref{eq:pqdu6} will yield $\svtotCsix$ rather than $\svtot$. Indeed, recent work studying $^{87}$Rb-Rb collisions which extracted the $C_6$ value from trap loss measurements, supports this assertion \cite{PhysRevA.106.052812}. In addition, the experimentally determined $\svtot$ values from \cite{Shen_2021} also support this interpretation, as illustrated in Table~\ref{tab:svtotexpC6}.
\begin{table}
\vspace{20pt}
\centering
\begin{tabular}{||c|c|c|c||}
\hline
& \multicolumn{3}{c||}{($10^{-15}$m$^3$/s)} \\
 Collision Pair &  $K(294\rm{K})$\cite{Klos:2023} & $\svtotCsix$ & $\svtot_{\rm{QDC}}$\cite{Shen_2021} \\
\hline
$^{87}$Rb-H$_2$ & 3.9(1)\phantom{2} & 4.98 & 5.12(15) \\
$^{87}$Rb-He &    2.37(3) & 2.49 & 2.41(14) \\
$^{87}$Rb-Ne &    2.0(2)\phantom{2} & 2.01 & --- \\
\hline
$^{87}$Rb-N$_2$ & 3.45(6) & 3.18 & 3.14(5) \\
$^{87}$Rb-Ar &    \phantom{2}3.031(7) & 2.81 & 2.79(5) \\
$^{87}$Rb-Kr &    2.79(1) & 2.63 & --- \\
$^{87}$Rb-Xe &    2.88(1) & 2.75 & 2.75(4) \\
\hline
$^{87}$Rb-Rb \cite{PhysRevA.106.052812} &  --- & 6.38 & \phantom{2}6.44(12) \\
\hline
\end{tabular}
\caption{\justifying{A comparison of the $\svtot$ values for $^{87}$Rb-X (X = H$_2$, He, Ne, N$_2$, Ar, Kr, and Xe) at T = 294K. The second column shows the FQMS values reported by Ref.\cite{Klos:2023}, $K(T) = K_0 + K_1(T-300\rm{K})$, the third shows the purely $C_6$ potential predictions, $\svtotCsix$, and the final column gives the values derived from fitting the universal model (Eq.~\ref{eq:pqdu6}) for trap loss to experimental measurements, $\svtot_{\rm{QDC}}$, reported in \cite{Shen_2021}. Also included is the result from \cite{PhysRevA.106.052812} for $^{87}$Rb-Rb collisions. The good agreement between the $\svtotCsix$ and the $\svtot_{\rm{QDC}}$ for heavy collision pairs is experimental evidence supporting the revised hypothesis that the trap loss rate variation with trap depth depends only on $\svtotCsix$ and not $\svtot$ as originally conjectured.}}
\label{tab:svtotexpC6}
\end{table}


Because of the presence of the long range $C_8$ and $C_{10}$ terms, which systematically increase the total rate coefficient, the purely experimental calibration procedure of extracting $U_d$ from measurements of trap loss rates will lead to an estimate of the rate coefficient that is systematically lower than the true value, $\svtotCsix < \svtot$. As discussed above, this discrepancy is up to 7\% (for $^{87}$Rb-Ar) collisions.  When the $C_6$, $C_8$, and $C_{10}$ coefficients are known \textit{a priori}, then the semi-classical approximation for the elastic scattering phase shifts offers an opportunity to determine a more accurate estimate of $\svtot$.

\section{Conclusions}

Using full quantum mechanical scattering calculations and realistic interaction potentials taken from Refs.~\cite{Klos:2023, Medvedev2018}, we have re-examined the original postulates of the universality hypothesis for room-temperature collisions reported in \cite{Booth2019, Shen_2020}.  We find evidence supporting the hypothesis that the total collision rate coefficient, $\svtot$, and the trap loss rate coefficient, $\svloss$, are universal for heavy collision partners with either a heavy ($^{87}$Rb) or a light ($^{7}$Li) sensor atom, determined solely by the long-range portion of the inter-species interaction potential.

This universality of $\svtot$ for room-temperature collisions is particularly significant to vacuum metrology because it implies that errors or uncertainties in the short-range part of the interaction potential do not propagate to the total collision rate coefficient and thus to the density or pressure inferred from a measurement.

For such universal collision partners, the total rate coefficient, $\svtot$, is well approximated by a semi-classical prediction, $\svtotSC$, that only accounts for the scattering phase shift arising from the long-range interaction potential terms, $C_6$, $C_8$, and $C_{10}$.  Specifically the SC values obtained in this work agree with the values computed by FQMS calculations using the complete potentials and published by K{\l}os Tiesinga \cite{Klos:2023} to better than 0.5\% for $^{87}$Rb-(N$_2$, Kr, Xe) collisions and to within 1.1\% for $^{87}$Rb-Ar. The agreement for $^{7}$Li-(N$_2$ and Ar) is within 0.5\%, and within 2\% for $^{7}$Li-(Kr and Xe). {The marginally larger discrepancy observed for collisions with $^{7}$Li sensor atoms may be attributed to two factors:  One contribution is the larger amplitude and lower frequency (in $v$ or $k = \mu v/\hbar$) glory oscillations in the total quantum mechanical cross-sections, $\sigma_{\rm{tot}}(v)$, associated with the smaller reduced mass for these pairs. For these cases, the averaging over the ambient temperature Maxwell-Boltzmann distribution of the background collisions inherent in the rate coefficient may not completely suppress the glory effects.  A second contribution to the discrepancy is the breakdown of the Jeffreys-Born phase approximation at small values of $L$.  Because the reduced mass of the collision complex is significantly smaller when using the $^{7}$Li sensor atom, far fewer partial waves contribute to the rate coefficient rendering the Jeffreys-Born approximation, and, thus, the SC approximation, less accurate.}



We also confirm the previously reported breakdown of this collision universality for light background gas collision partners.  The breakdown is due to the room temperature velocity average extending above $\vmaxone$ where the shape of $\sigma_{\rm{tot}}(v)$ is dominated by the shape of the interaction potential at short range.  Another feature of light background gas collision partners is that the shape of $\sigma_{\rm{tot}}(v)$ does not have a significant domain in relative speed where it follows the purely long-range character prediction.

Finally, we study the variation of the loss rate coefficient with trap depth for universal collision pairs and find that while $\svtot$ (equivalently $\svloss(U=0)$) depends on all of the long range terms ($C_6$, $C_8$, and $C_{10}$), the shape of the normalized trap loss rate coefficient ($\svloss(U)/\svtot = 1 - p(U)$), for small trap depths, only provides information about the leading order interaction potential term, $-C_6/r^6$. The consequence is that a fit of the experimentally measured loss rates normalized by the loss rate at zero trap depth will provide the energy scale $\Udsix$ and the corresponding $\svtotCsix$ instead of the total collision rate coefficient as originally postulated in \cite{Booth2019, Shen_2020}.
This finding together with the fact that the total collision rate coefficient for universal collision pairs is increased by the presence of the $C_8$ and $C_{10}$ long range terms implies that sensor atom self-calibration will provide an estimate, $\svtotCsix$, which is systematically below $\svtot$ by up to approximately 10\% (as shown in Tab.~\ref{tab:svtotexpC6}).
Finally, the expectation is that the variation of the normalized loss rate coefficient with trap depth for a non-universal collision pair will deviate from the universal function and will depend on the entire interaction potential.

We believe this work explains the observed discrepancies, reinforces the conjecture that room-temperature collisions are universal, insensitive to the short range interaction potential shape, demonstrates the simplicity of accurately estimating the rate coefficients for universal collision partners using the semi-classical approximations, and provides guidance for using atoms as a self-calibrating primary pressure standard.

\section{Acknowledgements}

\noindent{We acknowledge the financial support from the Natural Sciences and Engineering Research Council of Canada (NSERC grants RGPIN-2019-04200, RGPAS-2019-00055) and the Canadian Foundation for Innovation (CFI project 35724). This work was done at the Center for Research on Ultra-Cold Systems (CRUCS) and was supported in part, through computational resources and services provided by Advanced Research Computing at the University of British Columbia. K.W.M. acknowledges support from the Deutsche Forschungsgemeinschaft within the RTG 2717 program.}

\appendix
\section{Glory Oscillation Origins and Predictions}\label{theAppendix}
Universality relies on velocity averaging to minimize or eliminate the effects of the glory undulations inherent in the elastic scattering process. When successful, the averaging reveals the nature of the long-range potential while erasing any information about the short-range portion of the inter-species interaction. 

In this appendix we illustrate the origins of the glory undulations seen in the $\sigma_{\rm{tot}}(v)$ versus $v$ spectrum as an aid to define the limits of validity of the universality hypothesis. In particular, we use a Lennard-Jones (LJ) potential to derive the salient features of the glory oscillations. Namely,
\begin{enumerate}
\item We use the Jeffreys-Born approximation for the phase shifts to determine the partial wave which leads to glory scattering, $L_g$, for each $v$. 
\item Based on the analysis of Bernstein\cite{10.1063/1.1733383,10.1063/1.1733558} the upper limit of the speed over which the long-range portion of the potential dominates the $\sigma_{\rm{tot}}(v)$ values, $\vmaxone$, is estimated.
\item Using an expression from \cite{10.1063/1.1733558} the number of bound states the inter-species potential can support is computed and provides an estimate for the number of glory undulations observed over the cross-section spectrum \cite{child1996molecular, 10.1063/1.1733558} up to $v \approx \vmaxone$.
\item We compare the partial wave dependencies of the cross-sections for $^{87}$Rb-Ar to $^{7}$Li-Ar to demonstrate how the lighter trapped species leads to larger amplitude glory undulations.
\item Finally we provide an upper limit estimate of the maximum expected glory undulations-induced deviation to the values of the total collision coefficients, $\svtot$.
\end{enumerate}

For the rest of this appendix we will use a LJ potential to describe the inter-species collisions. While this is not a physically realistic potential, it allows us to derive tractable results which capture the behavior observed in FQMS calculations using more realistic potentials.

\subsection{Glory Oscillations}
\begin{center}
\begin{figure*}[t]
   \centering
    \begin{subfigure}[t]{0.48\textwidth}
        \centering
        \includegraphics[width=\linewidth]{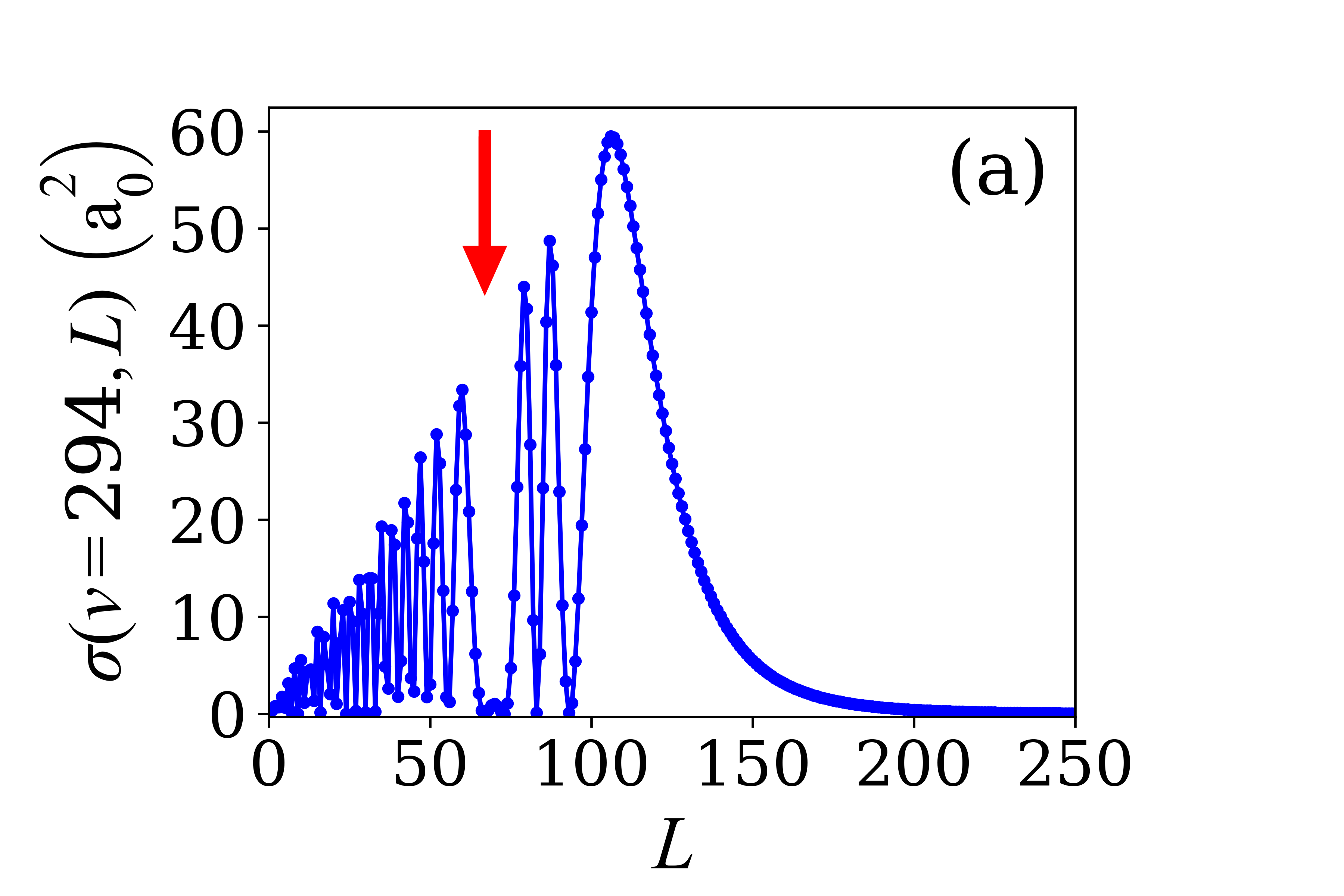} 
         \label{fig:RbArsig}
   \end{subfigure}
   \centering
       \begin{subfigure}[t]{0.48\textwidth}
       \centering
        \includegraphics[width=\linewidth]{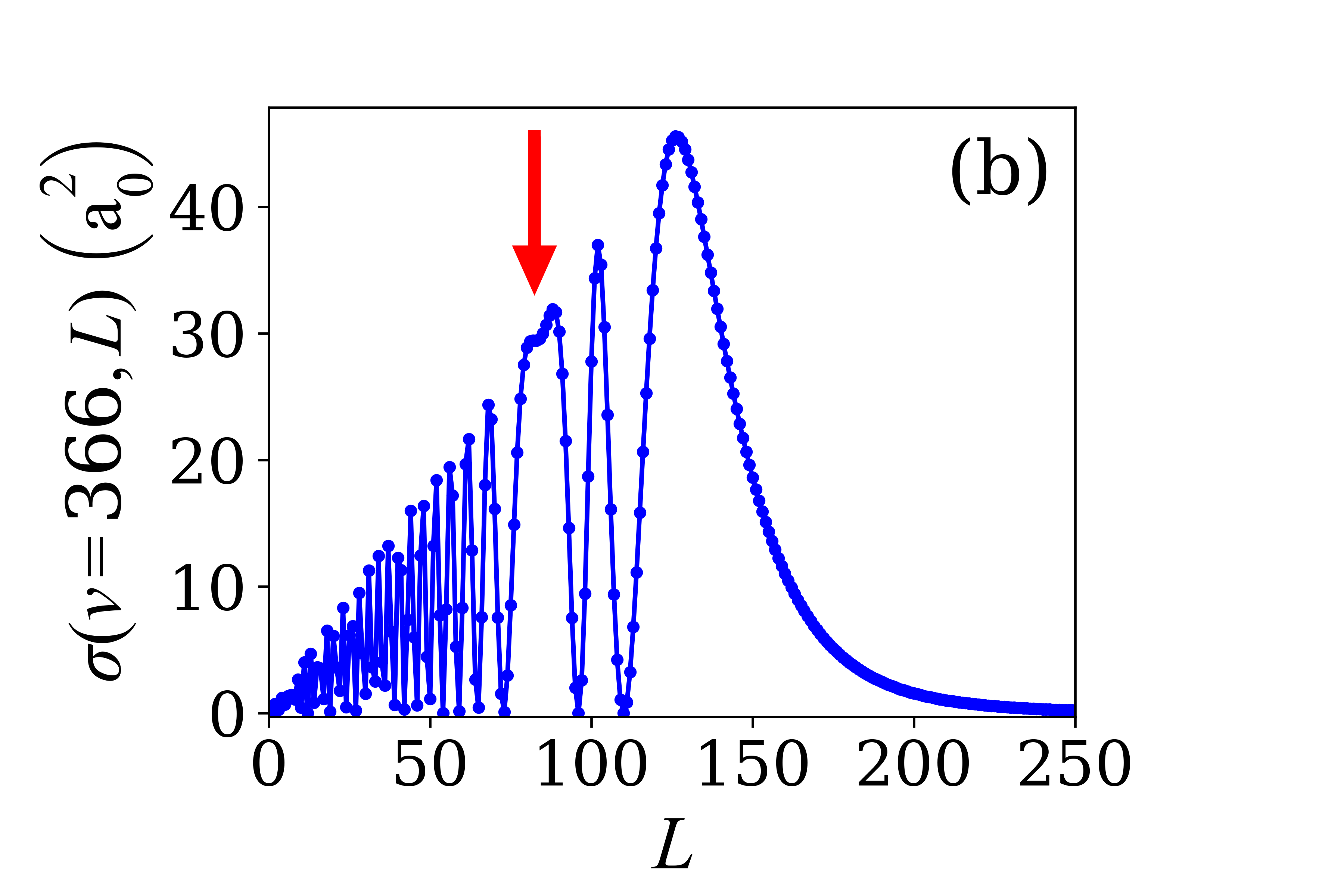} 
        \label{fig:RbHesig}
   \end{subfigure}

   \centering
    \begin{subfigure}[t]{0.48\textwidth}
        \centering
        \includegraphics[width=\linewidth]{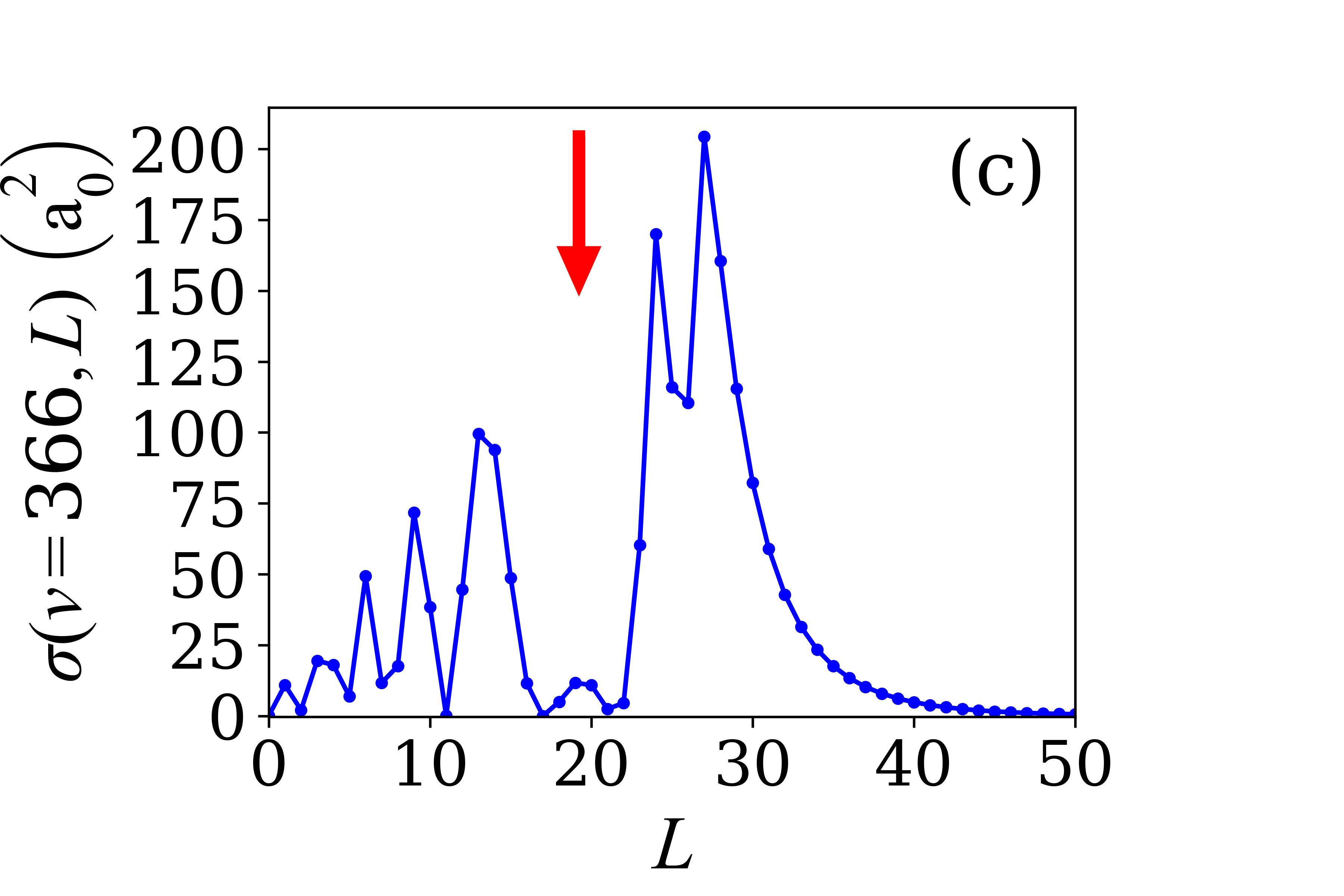} 
         \label{fig:RbArsig}
   \end{subfigure}
   \centering
       \begin{subfigure}[t]{0.48\textwidth}
       \centering
        \includegraphics[width=\linewidth]{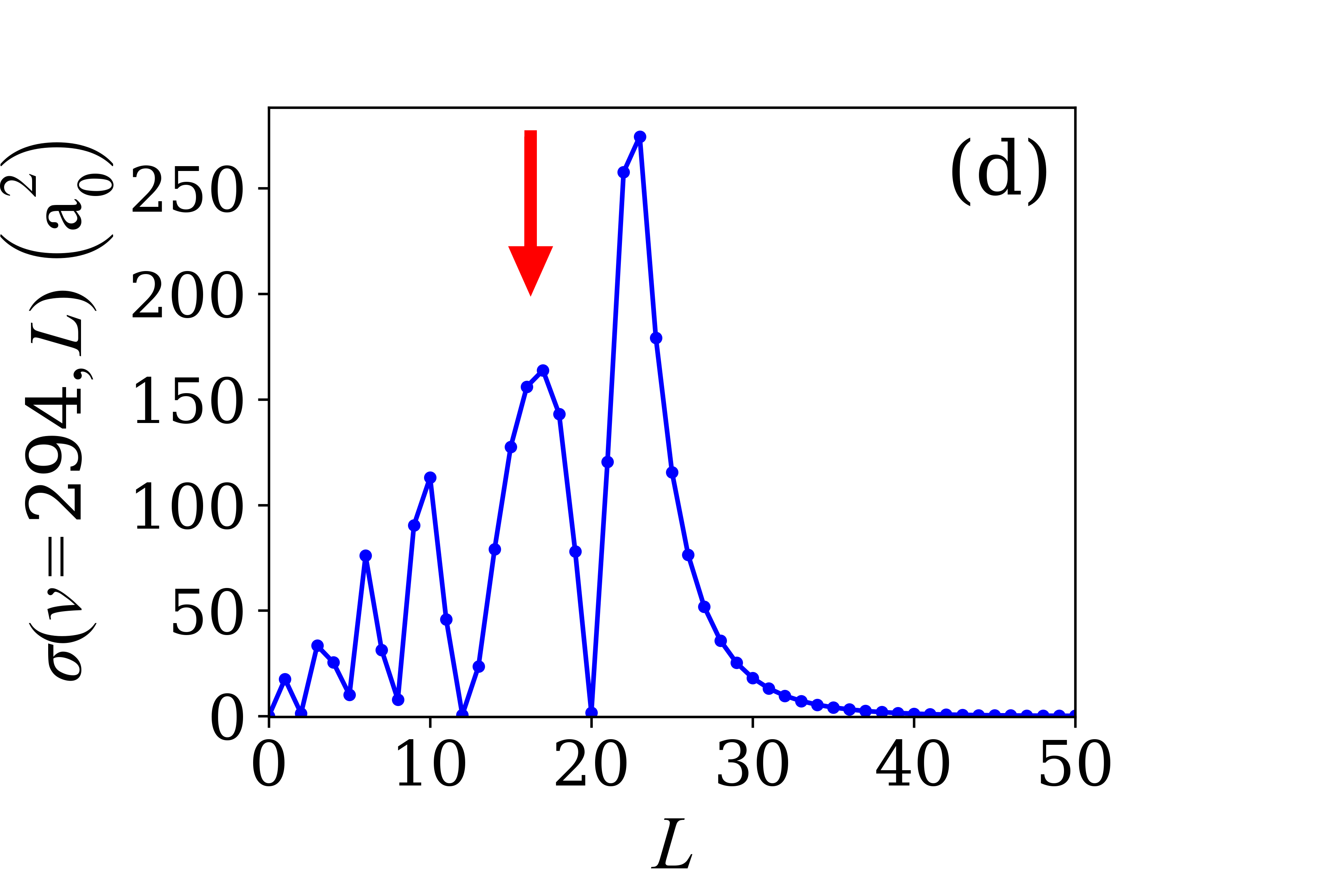} 
        \label{fig:RbHesig}
   \end{subfigure}
    \caption{\justifying{Illustrations of the effects of glory scattering interference: Plots of the $\sigma(v, L)$ versus $L$ for $^{87}$Rb-Ar using the LJ model (a) $v = 294$ m/s, (b) $v = 366$ m/s,  and using the K{\l}os Tiesinga (KT) potential \cite{Klos:2023} for $^{7}$Li-Ar (c) $v = 366$ m/s and (d) $v = 294$ m/s. The red arrow on each figure indicates the region near $L = L_g$ where variation of the scattering phase shifts in $L$ slows due to the short range contribution producing an interference minimum, as in (a) and (c), or a maximum, as in (b) and (d). This glory scattering interference changes from destructive to constructive as $v$ varies, leading to the undulations observed in Fig.~\ref{fig:RbArHeLJcompare}. }}
    \label{figA:RbArglory}
\end{figure*}
\end{center}

Glory undulations in the cross-section occur when lower~-~$L$ partial waves scatter in the forward direction ($\theta = 0$) and interfere with the large-$L$ partial waves which are also primarily scattered in the forward direction.  To illustrate this phenomenon, plots of $\sigma(v,L)$ versus $L$ for $^{87}$Rb-Ar collisions 
for a LJ potential are shown in Fig.~\ref{figA:RbArglory}(a) and (b) for collision speeds of 294 m/s and 366 m/s, respectively.
\bea 
\sigma(k,L) &=& \frac{4\pi}{k^2}\left(2L+1\right) \sin^2\eta_L(k)
\label{eq:AsigmakL}.
\eea
One observes that $\sigma(k, L)$ has a characteristic shape consisting of a lower-$L$ regime where it oscillates rapidly, corresponding to the rapid change in $\eta_L(k)$ (or, equivalently $\eta_L(v)$) with $L$. This is followed by a final lobe for which $0 \le \eta_L(k) \le \pi$. The peak of this lobe occurs at $\eta_L(k) \approx \pi/2$.

In the absence of glory scattering, the plots for different collision speeds would all look similar, with $\sigma(k, L)$ oscillating smoothly as the maxima grow monotonically following the $(2L+1)$ asymptote out to the final lobe. Glory scattering introduces interference which can reinforce or suppress the contributions of a small range of partial waves centered at the value $L_g$. Such interference is evident in the plots in Fig.~\ref{figA:RbArglory} with the approximate value of $L_g$ shown by the red arrows. (Fig.~\ref{figA:RbArglory}(a) demonstrates strong destructive interference, while (b) displays constructive interference.) The values of $\sigma_{\rm{tot}}(v)$ plotted in Figs.~\ref{fig:RbArHeLJcompare}, \ref{fig:RbArXeSCcompare} and \ref{fig:RbArXeNISTcompare}, are the sum of these $\sigma(k, L)$ over $L$. Thus, the glory undulations arise from the interference effects deviating the sum from the underlying, monotonic long-range prediction (e.g. Eq.~\ref{eq:svtot6JB} for a purely $C_6$ potential).

For elastic collisions \cite{child1996molecular} we can relate the scattering angle to the elastic scattering phase shift by,
\bea
\theta\left(v,L\right) &=& 2 \left(\frac{\partial \eta_L(v)}{\partial L}\right).
\label{eq:theta_eta}
\eea
For a LJ potential we can apply the Jeffreys-Born  (JB) approximation to write,
\bea 
\eta_L(k) &\approx& \frac{3\pi}{16}\left(\frac{\mu C_6}{\hbar^2} \right)\frac{k^4}{L^5} - \frac{63\pi}{512}\left(\frac{\mu C_{12}}{\hbar^2} \right)\frac{k^{10}}{L^{11}}.
\label{eq:etaLJJB}
\eea
(Here the $\ell = L + 1/2$ has been approximated by $L$.)
Setting Eq.~\ref{eq:theta_eta} equal to 0, and inserting Eq.~\ref{eq:etaLJJB}, one finds an estimate for the partial wave which leads to glory scattering, $L_g$,
\bea
L_g(k) &=& \left[\frac{21 \cdot 11}{32 \cdot 5} \right]^{\frac{1}{6}} k R_0 \ = \ 1.063 \left(\frac{\mu v}{\hbar}\right) R_0.
\label{eq:Lgk}
\eea
(Recall, $R_0 = \left(C_6/(4 D_e)\right)^{1/6}$ is the inter-species separation for which $V(R_0) = 0$.) This estimate works well for large $k = \mu v /\hbar$, for which there are a large number of partial waves making the JB approximation valid, but becomes less accurate for lower values of $k$ (or $v$).

\subsection{The Long-Range Dominated Region, $v < \vmaxone$}
Once the values of $L_g$ have been estimated, their corresponding Jeffreys-Born (JB) phases, $\eta_g(v)$, can be deduced from Eq.~\ref{eq:etaLJJB},
\bea 
\eta_g(v) &=& 0.1506 \cdot \frac{\pi}{2} \cdot \left( \frac{\vg}{v}\right)
\label{eq:etag}
\eea
where $\vg = 4 D_e R_0/\hbar$ (refered to as $v^*$ in Child \cite{child1996molecular}). It is evident that this approximation will be valid for larger $v$ and fails as $v \rightarrow 0$.

Bernstein \cite{10.1063/1.1733383,10.1063/1.1733558} studied the $v$ dependence of the $\eta_g(v)$ values to predict the relative collision speeds where the maxima and minima of the glory undulations in $\sigma_{\rm{tot}}(v)$ occur. He reported the maxima occur when $\eta_g = (N-3/8)\pi$ and the minima at $\eta_g = (N-3/8 + 1/2) \pi$ where $N = 1, 2, 3, \cdots$. The corresponding JB glory maxima will occur at velocities given by
\bea
v_{N}^{\rm{max}} &=& \frac{0.15062}{2\left(N-\frac{3}{8}\right)} \vg.
\label{eq:vNmax}
\eea
The value for $N = 1$ corresponds to the highest speed glory undulation maximum. This marks the upper relative collision speed limit where the long-range character of the potential dominates the $\sigma_{\rm{tot}}(v)$. We have labeled this value $\vmaxone = v_{N=1}^{\rm{max}}$, Eq.~\ref{eq:vmax1}, in the main body of this work. $\vmaxone$ is a good estimate for the LJ potential since the JB approximation for the phase shifts is valid for large $v$. Indeed, the $\vmaxone$ is also a good estimate for the location of the final maximum in $\sigma_{\rm{tot}}(v)$ for more realistic potentials such as those of references \cite{Klos:2023, Medvedev2018}. Table~\ref{tab:LJKTcomparisons} provides a comparison of the values for $R_0$, $\vg$, $\vmaxone$, and $N_0$ (the number of rotation-less bound states supported by the potential) for the K{\l}os and Tiesinga (KT) published potentials \cite{Klos:2023} and the analytically derived values for a LJ potential with the same $C_6$ and potential depth, $D_e$.  These values are compared for both $^{7}$Li-X and $^{87}$Ar-X (X = H$_2$, He, Ne, N$_2$, Ar, Kr, and Xe) collisions. In particular we observe that the LJ $\vmaxone$ values agree with the KT model values within 5\% for the heavy background collision partners. Thus the LJ based value is a good estimate of the high speed glory maximum for the FQMS computation.

\begin{center}
\begin{table*}
\centering
\begin{tabular}{||c|c|cc|c|c|cc|ccc|cc||}
\hline
  &\phantom{ }$\mu$ (amu)\phantom{ }  &\multicolumn{2}{c|}{$R_0$ ($\rm{a_0}$)} &\phantom{ }  $\vp$ (m/s) \phantom{ }& \phantom{ } $\vdvmax$ (m/s) \phantom{ }&\multicolumn{2}{c|}{$\vg$ (m/s)} & \multicolumn{3}{c|}{$\vmaxone$ (m/s)} & \multicolumn{2}{c||} {$N_0$} \\
 Collision Pair &  &\phantom{12}  LJ \phantom{12}& \phantom{12} KT \phantom{12} &  &  &\phantom{12} LJ \phantom{12}& \phantom{12}KT \phantom{12} & \phantom{12} LJ \phantom{12} &\phantom{12} KT \phantom{12}&\phantom{ } KT:LJ \phantom{ }&\phantom{}  LJ \phantom{ }& \phantom{ }KT \phantom{ }\\
\hline
$^{87}$Rb-H$_2$ & 1.97 & 9.73 & 10.22 & 1558 & 1908 & 3725 & 3912 & 449 & 472 & 1.05 & 2 & 2\\
$^{87}$Rb-He & 3.83 &  11.34 & 12.20 & 1105 & 1353 & 515 & 555 & 62 & 67 & 1.08 & 1 & 1\\
$^{87}$Rb-Ne & 16.38 & 9.59 & 10.04 & 492 & 603 & 2332 & 2441 & 281 & 294 &  1.05 & 3 & 4\\
\hline
$^{87}$Rb-N$_2$ & 21.18 & 9.67 & 10.07 & 418 & 512 & 9022 & 9396 & 1087 & 1132 & 1.04 & 8 & 8\\
$^{87}$Rb-Ar & 27.37 & 8.83 & 8.79 &  350 & 429 & 13593 & 13531 & 1638 & 1630 & 0.995 & 10 & 10\\
$^{87}$Rb-Kr &  42.66 & 8.59 & 8.29 & 242 & 296 & 23117 & 22998  & 2785 & 2687 & 0.965 & 16 & 16\\
$^{87}$Rb-Xe & 52.29 & 8.64 & 8.27 & 193 & 236 & 35216 & 33697 & 4243 & 4060 & 0.957 & 22 & 21\\
\hline
\hline
$^{7}$Li-H$_2$ & 1.57 & 8.37 & 8.58 & 1558 & 1908 & 4405 & 4516 & 531 & 544 & 1.03 & 1 & 1\\
$^{7}$Li-He &  2.55 & 9.52 & 10.14 & 1105 & 1353 & 627 & 667 & 76 & 80 & 1.06 & 1 & 1\\
$^{7}$Li-Ne & 5.21 & 8.30 & 8.67 & 492 & 603 & 2417 & 2524 & 291 & 304 & 1.04 & 2 & 2\\
\hline
$^{7}$Li-N$_2$ & 5.61 & 8.46 & 8.78 & 418 & 512 & 9248 & 9593 & 1114 & 1156 & 1.04 & 4 & 4\\
$^{7}$Li-Ar &  5.97 & 7.87 & 7.95 & 350 & 429 & 12523 & 12647 & 1509 & 1524 & 1.01 & 4 & 4\\
$^{7}$Li-Kr &  6.47 & 7.75 & 7.69 & 242 & 296 & 20270 & 20107 & 2442 & 2423 & 0.992 & 6 & 6\\
$^{7}$Li-Xe & 6.66 & 7.81 & 7.65 & 193 & 236 & 30796 & 30177 & 3711 & 3636 & 0.980 & 7 & 7\\
\hline
\end{tabular}
\caption{\justifying{A comparison of the $\mu$, $R_0$, $\vg$, $\vmaxone$ and $N_0$ values for the K{\l}os and Tiesinga (KT) potential \cite{Klos:2023} and the analytically dervied values for a Lennard-Jones (LJ) potential with the same $C_6$ and $D_e$.  Also included is the ratio of the KT:LJ $\vmaxone$ (column 9) and the Maxwell-Boltzmann distribution peak speeds, $\vp$, at T = 294K (column 4).} }
\label{tab:LJKTcomparisons}
\end{table*}
\end{center}

\subsection{The number of glory undulations in $\sigma_{\rm{tot}}(v)$ vs $v$ : $N_0$}
Bernstein  \cite{10.1063/1.1733383,10.1063/1.1733558} and Child \cite{child1996molecular} assert that the number of glory undulations that appear in the $\sigma_{\rm{tot}}(v)$ versus $v$ spectrum is equal to the number of rotation-less bound states that can be supported by the potential, $N_0$. For a LJ potential this value is \cite{10.1063/1.1733558}
\bea
N_0 &=& 0.27\sqrt{\frac{2 \mu D_e R_0^2}{\hbar^2}} + \frac{1}{2}.
\label{eq:N0}
\eea
LJ potentials are completely defined by two parameters, $(C_6, C_{12})$ or $(D_e, R_0)$, related by,
\bea
C_6 &=& 4 D_e R_0^6 \\
C_{12} &=& 4 D_e R_0^{12}.
\eea
Here it is convenient to define the LJ potentials with $C_6$ and $D_e$ which allows $N_0$ to be expressed as,
\bea
N_0 &=& 0.27\sqrt{\frac{2 \mu}{\hbar^2}} \left[\frac{C_6 D_e^2}{4}\right]^{\frac{1}{6}}  + \frac{1}{2}.
\label{eq:N02}
\eea
From this expression one observes the number of bound states depends on the reduced mass of the colliding partners, the depth of the potential, and on the $C_6$ value. Table~\ref{tab:LJKTcomparisons} lists the values of $N_0$ determined using a LJ potential and using the KT potentials \cite{Klos:2023}. We observe that they agree within 1 for the two types of potentials.

\subsection{Estimate of the Relative Amplitude of Glory Undulations}
We observe that the amplitude of the glory scattering induced undulation super-imposed on the underlying long-range character of the $\sigma_{\rm{tot}}(v)$ versus $v$ has the following properties (See Fig.~\ref{fig:RbArXeNISTcompare}).
\begin{enumerate}
    \item The amplitude of the glory undulations is larger for lower reduced mass collision partners.
    \item The amplitude of the glory undulations reduces as the relative collision speed increases for each collision pair. 
\end{enumerate}
Here we will provide an explanation for these two observations and estimate the undulation amplitude to demonstrate its dependence on the reduced mass, $\mu$, $C_6$, and $D_e$. 

The effects of the interference generated by glory scattering is shown in Fig.~\ref{figA:RbArglory}:  Each of the plots of $\sigma(v, L)$ vs $L$ share a similar character. From, Eq.~\ref{eq:AsigmakL} each partial wave contribution, $\sigma(v,L)$ has a $(2L+1) \sin^2\eta_L(v)$ dependence.
The $\eta_L(v)$ phase shifts rise monotonically from a negative value at $L = 0$ (corresponding to the phase induced by the repulsive core \cite{child1996molecular}) through zero, reaching a positive peak at $L = L_g$, and then decreasing monotonically to zero as $L$ increases. 
Thus, the $\sigma(v, L)$ values oscillate for lower $L$ values  and increase in amplitude with $L$. This oscillation ends in a final ''lobe" where $\sigma(v, L_{\pi}) = 0$ corresponding to $\eta_{L_{\pi}}(v) = \pi$ rising to the final peak appearing near $\eta_{L_{\pi/2}}(v) = \pi/2$, and then decreasing monotonically towards zero as $L \rightarrow \infty$ corresponding to $\eta_L(v) \rightarrow 0$. Using the Jeffreys-Born approximation for a purely $C_6$ long-range potential the position of $L_{\pi}$ is,
\bea
L_{\pi}(k) &\approx& \left[\left(\frac{3}{16}\right) \left(\frac{\mu C_6}{\hbar^2 k}\right) \right]^{\frac{1}{5}} k \nonumber \\
L_{\pi}(v)&\approx& \left[\left(\frac{3}{16}\right) \left(\frac{ C_6}{\hbar}\right) \right]^{\frac{1}{5}} \cdot \left(\frac{\mu}{\hbar}\right) \cdot v^{\frac{4}{5}}
\label{eq:ALpi}.
\eea
This estimate indicates that the number of partial waves included in each $\sigma_{\rm{tot}}(v)$ increases as $\mu$, as $C_6^{\frac{1}{5}}$ and as $v^{\frac{4}{5}}$.  Thus, one observes that the $^{87}$Rb-Ar examples shown in Fig.~\ref{figA:RbArglory} encompass many more partial waves than the corresponding $^{7}$Li-Ar values.

For $v < \vmaxone$ these glory interference effects will occur with $L_g < L_{\pi}$. Thus, the glory-induced variation in the $\sigma(v)$ value can be approximated as the area of the slice of size $\Delta L$ centered on $L_g$,
\bea
\Delta \sigma_{g}(v) & \approx & \frac{2\pi}{k^2} L_g \cdot \Delta L \nonumber \\
&\approx& \frac{2\pi}{k^2} \left[\frac{5\cdot 32}{11 \cdot 21} \right]^{\frac{1}{5}} (k R_0) \cdot \Delta L \nonumber \\
&\approx& 2 \pi\left[\frac{5\cdot 32}{11 \cdot 21} \right]^{\frac{1}{5}} \left( \frac{\hbar}{\mu }\right)\cdot\left(\frac{C_6}{4 D_e}\right)^{\frac{1}{6}} \cdot \frac{\Delta L}{v}
\label{eq:AdeltaSigma}
\eea
This value can be compared to the purely long-range $C_6$ estimate,
\bea 
\sigma_{\rm{tot}}(v) & \approx & 8.0828 \left(\frac{C_6}{\hbar v}\right)^{\frac{2}{5}}
\label{eq:AsigmaC6}
\eea
to provide an estimate of the size of the relative glory undulation amplitude,
\bea
\frac{\Delta \sigma_{g}(v)}{\sigma_{\rm{tot}}(v)} & \approx & 0.722  \left[\frac{\hbar}{C_6}\right]^{\frac{2}{5}} \left(\frac{\hbar}{\mu}\right) \frac{1}{v^{\frac{3}{5}}}\cdot R_0 \cdot \Delta L 
\label{eqA:FractUndulation}
\eea
\newline
This estimate demonstrates that the glory undulation amplitude scales as $1/\mu$, leading to larger oscillations for lighter reduced mass. Assuming the $R_0$ and $\Delta L$ values are approximately equal for $^{7}$Li-Ar and $^{87}$Rb-Ar, Eq.~\ref{eqA:FractUndulation} predicts that the glory undulations in $^{7}$Li-Ar will be about 5 times larger than for $^{87}$Rb-Ar. 

Finally, the $1/v$ dependence in Eq.~\ref{eq:AdeltaSigma} is consistent with the observation that the glory undulation amplitude decreases with increasing relative collision speed. (See Fig.~\ref{fig:RbArXeLiArXecompare}(c), for example.)

\begin{figure}[h]
   \centering
    \includegraphics[width=0.48\textwidth]{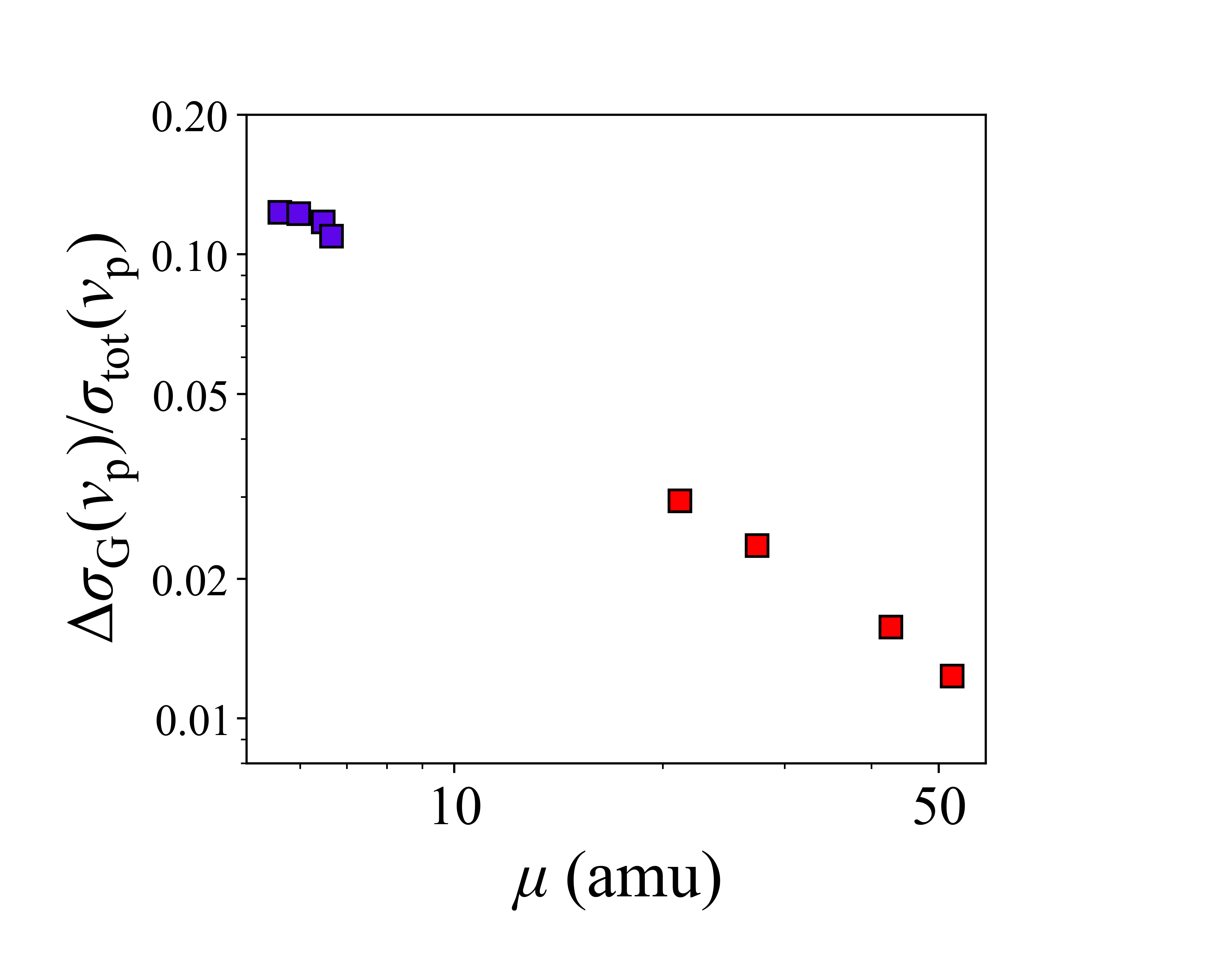}
    \caption{\justifying{Plot of $\/\Delta \sigma_{\rm{G}}(\vp)/\sigma_{\rm{tot}}(\vp)$ (for T = 294K) versus $\mu$ using $v = \vp$ as an estimate of the maximum residual effect of the glory undulations on the $\svtot$ values. Here the $^{87}$Rb-X  and $^{7}$Li-X where X = N$_2$, Ar, Kr, and Xe are the species expected to be universal for which Eq.~\ref{eqA:FractUndulation} is applicable. The $^{87}$Rb-X are the red squares, while the violet squares are the $^{7}$Li-X values. Here $\Delta L = 10$ was used for these estimates.}}
    \label{figA:LiRbUndulationsAmplitude}
\end{figure}

Eq.~\ref{eqA:FractUndulation} with v = $\vp$ can be used to estimate the upper bound on the amplitude of the glory undulations to the underlying long-range (universal) prediction for $\svtot$. That is, the ratio of $\Delta \sigma_{\rm{g}}(\vp) / \sigma_{\rm{tot}}(v)$ provides an estimate of the maximum relative residual effect of the glory oscillations on the total collision coefficients, $\svtot$. In the absence of the velocity averaging minimization of the glory oscillations, this is the maximum glory scattering induced deviation from the long-range background cross-section one would expect. These values are shown on Fig.~\ref{figA:LiRbUndulationsAmplitude} and in Table~\ref{tabA:LiRbUndulationsAmplitude}. As expected, the lower the reduced mass of the collision partners, the larger the potential for the glory undulations to affect the value of $\svtot$. 

The actual residual effect of the glory undulations on the total collision rate coefficients, $\svtot$, will depend on the number of undulations captured under the $v\cdot d(v)$ distribution used for the velocity averaging.

\begin{table}[h]
\centering
\begin{tabular}{||c|c|c||}
\hline
\multicolumn{3}{||c||}{$\Delta \sigma_{\rm{g}}(\vp)/\sigma_{\rm{tot}}(\vp)$}\\
\hline
{Background} & \multicolumn{2}{c||}{Sensor}\\
 Gas & $^{87}$Rb & $^{7}$Li \\
\hline
N$_2$ &\phantom{ } 0.029 \phantom{ }& \phantom{ } 0.123 \phantom{ }\\
Ar &  \phantom{ }  0.024 \phantom{ } & \phantom{ } 0.122  \phantom{ }\\
Kr &  \phantom{ }  0.016 \phantom{ } & \phantom{ } 0.117 \phantom{ }\\
Xe &  \phantom{ }  0.012 \phantom{ } & \phantom{ } 0.109 \phantom{ }\\
\hline
\end{tabular}
\caption{\justifying{Estimates of the maximum relative deviation of the $\svtot$ values from the $\svtotSC$ due to the effects of glory undulations at T = 294K. The values are shown for $^{87}$Rb-X (X = N$_2$, Ar, Kr, and Xe), column 2, and for $^{7}$Li-X (column 3). An estimate of $\Delta L = 10$ was used for these computations.}}
\label{tabA:LiRbUndulationsAmplitude}
\end{table}

\section{The Jeffreys-Born (JB) Approximation}

Child \cite{child1996molecular} describes the Jeffreys-Born (JB) approximation for estimating the large $L$ elastic collision phase shifts, $\eta_L(k)$ in Chapter 4. This relies on the Born approximation for the wavefunctions describing the collision. Namely, the zero order approximation is
\bea 
\psi^0_{L}(r) &\approx& k r j_L(kr)
\label{eqB:psi0}
\eea
where $j_L(kr)$ is the Bessel function of order $L$.
Using Green's functions, Child then derives, 
\bea
\eta_L(k) & = & -\frac{1}{k}\int_0^{\infty} \frac{2\mu V(r)}{\hbar^2} \left[\psi^0_{L}(r)\right]^2 dr.
\label{eqB:Childeta}
\eea
Different approximations for the expression for the wavefunctions, $\psi^0_{L}(r)$, lead to different approximations for $\eta_L(k)$. 

\subsection{The Jeffreys-Born Approximation, JB$_0$.}
The initial approximation, JB$_0$, for the phase shifts, is to use the large $r$ asymptotic form for the Bessel functions to determine the phase shifts. Namely,
\bea
\left[\psi^0_{L}(r)\right]^2 & \approx & k^2r^2j_L(kr)^2 \nonumber \\
 &\rightarrow& \frac{1}{2\sqrt{1-\frac{L^2}{k^2 r^2}}}.
\label{eqB:psi0squared}
\eea
Using Eq.~\ref{eqB:psi0squared} in Eq.~\ref{eqB:Childeta}, with $V(r) = -C_6/r^6$ one obtains,
\bea
\eta(k,L) &\approx&  \frac{1}{k}\int_0^{\infty} \frac{2\mu C_6}{\hbar^2 r^6} \left[\psi^0_{L}(r)\right]^2  dr \nonumber \\
&\approx& \frac{2\mu C_6}{\hbar^2 k}\int_{L/k}^{\infty} 
\frac{dr}{2 r^6 \sqrt{1-\left(\frac{L}{kr}\right)^2}} \nonumber \\
&\approx &\left(\frac{\mu C_6}{\hbar^2}\right) \frac{1}{k} \int_{L/k}^{\infty} \frac{dr}{r^6 \frac{L}{kr}\sqrt{\left(\frac{kr}{L}\right)^2-1}} \nonumber \\
&\approx& \left( \frac{\mu C_6} {\hbar^2}\right) \frac{k^4}{L^5} \int_0^{1} \frac{dz}{z^5 \sqrt{z^2-1}}
\label{eqB:etaJB61}
\eea
where $z = kr/L$.
This can be generalized for any $V(r)~=~C_n/r^n$,
\bea
\eta_L(k, C_n, \rm{JB}_0) & \approx & -\left( \frac{\mu C_n} {\hbar^2}\right) \frac{k^{n-2}}{L^{n-1}} \int_0^{1} \frac{dz}{z^{n-1} \sqrt{z^2-1}}.
\label{eqB:etaJBCn1}
\eea
The integral over $z$ gives 
\bea
\int_0^{1} \frac{dz}{z^{n-1} \sqrt{z^2-1}} &=& \sqrt{\pi} \frac{\Gamma\left(\frac{1}{2}\left(n-1\right)\right)}{\Gamma\left(\frac{n}{2} \right)}
\label{eqB:intz}
\eea
providing the prefactor in front of each phase shift. This approximation has been used to deduce the values of $L_g$ and $L_{\pi}$ in Appendix A.

\subsection{Refining the approximation, JB$_1$.}
Child continues his discussion of the phase shifts by making the next approximation using the Bessel function form of the wavefunction in Eq.~\ref{eqB:Childeta},
\bea
\eta_L(k) &\approx&  -\frac{1}{k}\int_0^{\infty} \frac{2\mu V(r)}{\hbar^2} \left[k^2 r^2 j_L(kr)^2\right] dr.
\eea
Child \cite{child1996molecular} reports the result for a potential, $V(r) = C_n/r^n$, which is labeled JB$_2$ here,
\bea
\eta_L(k, C_n, \rm{JB}_2) &\approx& -\frac{2\mu k C_n}{\hbar^2 } \int_{0}^{infty} \frac{1}{r^{n-2}} j_L(kr)^2 dr \nonumber \\
&=& -\left(\frac{\pi }{2^{n-1}}\right) \left[\frac{(n-2)! }{(\frac{n}{2}! \cdot \frac{2}{n})^2}\right]  \left(\frac{\mu C_n}{\hbar^2}\right) \cdot k^{n-2}  \cdot q(L,n) \nonumber \\
\label{eqB:etaJB2}
\eea
where,
\bea
q(L,n) &=&  \frac{\Gamma(L + \frac{1}{2} - \frac{n}{2} + 1)}{\Gamma(L + \frac{1}{2} - \frac{n}{2} + n)}.
\label{eqB:qLn}
\eea
Using the properties of Gamma functions, the ratio can be simplified, leaving an expression in the denominator which is the product of $n-1$ terms of the form $L + 1/2 - n/2 + m$ where $m \in [n-1, n-2, ..., 1]$. The resulting polynomial has the leading term $L^{n-1}$. 
For large, $L$, only the leading term in the denominator is significant and $q(n,L) \rightarrow 1/L^{n-1}$ so that  Eq.~\ref{eqB:etaJB2} reduces to the JB$_0$ estimate, Eq.~\ref{eqB:etaJBCn1}.

Replacing $\ell = L + 1/2$, one has,
\bea
q(L,n) &=& \frac{1}{\left(\ell  - \frac{n}{2} + (n-1) \right)\left(\ell - \frac{n}{2} + (n-2) \right)\cdots \left(\ell - \frac{n}{2} + 1 \right)} \nonumber \\
&=& \frac{1}{\ell^{n-1}\left(1 + \frac{n-1 + \frac{n}{2}}{\ell} \right)\left(1 + \frac{n-2 + \frac{n}{2}}{\ell} \right)\cdots \left(1 + \frac{1 + \frac{n}{2}}{\ell} \right)}. \nonumber \\
\label{eqB:ellfull}
\eea
Again, for large $\ell$, one has,
\bea
q(L,n, \rm{JB}_1) &\approx&\frac{1}{\ell^{n-1}}
\label{eqB:ellJB1}
\eea
and one obtains the JB$_1$ approximations applied in the main body of this paper,
\bea
\eta_L(k, C_n, \rm{JB}_1) &\approx& -\left(\frac{\pi }{2^{n-1}}\right) \left[\frac{(n-2)! }{(\frac{n}{2}! \cdot \frac{2}{n})^2}\right]  \left(\frac{\mu C_n}{\hbar^2}\right) \cdot \frac{k^{n-2}}{\ell^{n-1}}. \nonumber \\
\label{eqB:etaJB1}
\eea
\begin{table}[t]
\centering
\begin{tabular}{|r|cc|c|ccc||}
\hline 
 & \multicolumn{2}{|c|}{FQMS $\svtot$}& $K(294 \rm{K})$ &\multicolumn{3}{|c||}{$\svtotSC$}  \\
  & \multicolumn{2}{|c|}{($10^{-15}$m$^3$/s)}& ($10^{-15}$m$^3$/s) &\multicolumn{3}{|c||}{($10^{-15}$m$^3$/s)} \\
 &  KT\cite{Klos:2023} & Med\cite{Medvedev2018} &   \cite{Klos:2023} &  \phantom{2}JB$_2$ &  JB$_1$ & JB$_0$\phantom{2}\\  
\hline
 $^7$Li-H$_2$ &      3.11 &     --- &           3.16(6) &        \phantom{2}4.48 &       4.46 &        4.61\phantom{2} \\
 $^7$Li-He &      1.65 &         --- &       1.65(4) &        \phantom{2}2.31 &       2.31 &        2.37\phantom{2} \\
 $^7$Li-Ne &      1.56 &     --- &          \phantom{2}1.56(14) &        \phantom{2}1.72 &       1.71 &        1.76\phantom{2} \\
 $^7$Li-N$_2$ &      2.60 &     --- &           2.63(2) &        \phantom{2}2.62 &       2.63 &        2.70\phantom{2} \\
 $^7$Li-Ar &      2.33 &     --- &           \phantom{2}2.329(6) &        \phantom{2}2.30 &       2.32 &        2.35\phantom{2} \\
 $^7$Li-Kr &      2.14 &     --- &           \phantom{2}2.145(4) &        \phantom{2}2.13 &       2.11 &        2.22\phantom{2} \\
 $^7$Li-Xe &      2.24 &     --- &           2.24(2) &        \phantom{2}2.21 &       2.21 &    2.30\phantom{2} \\
 \hline
 $^{87}$Rb-H$_2$ &      3.93 &     --- &           \phantom{2}3.88(10) &        \phantom{2}5.92 &       5.89 &        6.03\phantom{2} \\
 $^{87}$Rb-He &      2.37 &     2.44 &           2.37(3) &        \phantom{2}3.09 &       3.08 &        3.14\phantom{2} \\
 $^{87}$Rb-Ne &      2.04 &     1.98 &           \phantom{2}2.00(20) &        \phantom{2}2.27 &       2.28 &        2.29\phantom{2} \\
 $^{87}$Rb-N$_2$ &      3.44 &     --- &           3.45(6) &        \phantom{2}3.45 &       3.45 &        3.48\phantom{2} \\
 $^{87}$Rb-Ar &      3.02 &     3.04 &           \phantom{2}3.031(7) &        \phantom{2}3.01 &       3.00 &        3.02\phantom{2} \\
 $^{87}$Rb-Kr &      2.78 &     2.79 &           2.78(1) &        \phantom{2}2.77 &       2.77 &        2.78\phantom{2} \\
 $^{87}$Rb-Xe &      2.87 &     2.88 &           2.87(1) &        \phantom{2}2.88 &       2.88 &        2.87\phantom{2} \\
\hline
\end{tabular}
\caption{\justifying{$^7$Li-X and $^{87}$Rb-X (X = H$_2$, He, Ne, N$_2$, Ar, Kr, and Xe) $\svtot$ values from the FQMS based on K{\l}os and Tiesinga (KT) potential \cite{Klos:2023}, the Medvedev \textit{et al} (Med) potential \cite{Medvedev2018}, the $K(T) = K_0 + K_1(T - 300\rm{K})$ values from \cite{Klos:2023}, and our SC computations $\svtotSC$ using the different JB approximations for $\eta_L(k)$. The background gas ensemble temperature is 294 K.}}
\label{tabB:RbLiX}
\end{table}

For a purely attractive $C_6$ potential ($V(R) = -C_6/r^6$), the $\svtotCsix$ corresponding to JB$_0$ and JB$_1$ can be written analytically:
\bea
\svtotCsix^{JB_0} &\approx& 8.49463 \left(\frac{C_6}{\hbar \vp}\right)^{\frac{2}{5}}\vp + 7.19729 \left(\frac{\hbar}{\mu}\right)\left(\frac{C_6}{\hbar \vp}\right)^{\frac{1}{5}}\nonumber \\
\label{eqB:svtotC6JB0}
\eea
and
\bea
\svtotCsix^{JB_1} &\approx& 8.49463 \left(\frac{C_6}{\hbar \vp}\right)^{\frac{2}{5}}\vp.
\label{eqB:svtotC6JB1}
\eea
One notes the JB$_0$ approximation contains an extra term in $\hbar/\mu$. The JB$_2$ approximation for $\eta_L(k)$ (Eq.~\ref{eqB:etaJB2}) does not yield a simple analytical expression but can be used to compute $\svtotCsix^{JB_2}$ numerically.

To evaluate the differences between the approximations, JB$_0$, JB$_1$, and JB$_2$, the $\svtotSC$ were computed using the three approximations for $\eta_L(k)$. These were compared to FQMS computations based on the K{\l}os and Tiesinga (KT) \cite{Klos:2023} and Medvedev \textit{et al} (Med) \cite{Medvedev2018} potentials, and to the values of $K(T) = K_0 + (T-300\rm{K})$ reported by K{\l}os and Tiesinga \cite{Klos:2023}. (See Table~\ref{tabB:RbLiX}.) The background gas temperature used in these computations was 294 K.  

In addition, Table~\ref{tabB:RbLiXC6} contains the $\svtotCsix$ values computed based on the analytical JB$_0$ approximation (a-JB$_0$), Eq.~\ref{eqB:svtotC6JB0},  the analytical JB$_1$ approximation (a-JB$_1$), Eq.~\ref{eqB:svtotC6JB1}, and the numerically computed $\svtotCsix$ values, JB$_0$, JB$_2$, and JB$_1$.

\begin{table}[t]
\vspace{0pt}
\centering
\begin{tabular}{||r|ccccc||}
\hline 
 & \multicolumn{5}{|c||}{$\svtotCsix$} \\
   & \multicolumn{5}{|c||}{($10^{-15}$m$^3$/s)}\\
 &  a-JB$_0$  &  a-JB$_1$ &  JB$_0$ & JB$_1$  &  JB$_2$ \\
\hline
 $^7$Li-H$_2$ & 
4.09 &      3.94 &       4.09 &   3.94 &       3.95\phantom{2}  \\
 $^7$Li-He &   1.98 &      1.90 &       1.98 &   1.90       & 1.91\phantom{2} \\
 $^7$Li-Ne &      1.58 &      1.53 &       1.58 &           1.52 &       1.52\phantom{2} \\
 $^7$Li-N2 &      2.53 &      2.46 &       2.54 &          2.47 &  2.45\phantom{2} \\
 $^7$Li-Ar &   2.22 &      2.16 &       2.22 &           2.17 & 2.16\phantom{2}        \\
 $^7$Li-Kr &   2.10 &      2.03 &       2.11 &         2.04 &  2.04\phantom{2} \\
 $^7$Li-Xe &    2.21 &      2.13 &       2.25 &         2.14 &  2.14\phantom{2} \\
\hline
  $^{87}$Rb-H$_2$ &    5.12 &      4.98 &       5.11 &   4.98 &         5.00\phantom{2}  \\
 $^{87}$Rb-He &    2.55 &      2.49 &       2.55 &       2.49 &    2.49\phantom{2}\\
 $^{87}$Rb-Ne &    2.03 &      2.01 &       2.03 &       2.00 &    2.01\phantom{2} \\
 $^{87}$Rb-N$_2$ &    3.20 &      3.18 &       3.20 &      3.18 &     3.18\phantom{2} \\
 $^{87}$Rb-Ar &    2.82 &      2.81 &       2.82 &        2.82 &   2.81\phantom{2} \\
 $^{87}$Rb-Kr &    2.64 &      2.63 &       2.64 &       2.63    &  2.63\phantom{2} \\
 $^{87}$Rb-Xe &    2.76 &      2.75 &       2.77 &       2.75 &     2.75\phantom{2} \\
\hline
\end{tabular}
\caption{\justifying{The $\svtotCsix$ values of $^7$Li-X and $^{87}$Rb-X $\svtotCsix$ for X = (H$_2$, He, Ne, N$_2$, Ar, Kr, and Xe) based on the analytical JB$_0$ (a-JB$_0$) expression, Eq.~\ref{eqB:svtotC6JB0}, the analytical expressions JB$_1$ (a-JB$_1$), Eq.~\ref{eqB:svtotC6JB1}, and substantiated by numerical computations for the JB$_0$, JB$_1$, and JB$_2$ approximations of the phase shifts $\eta_L(k)$. Here T = 294 K.}}
\label{tabB:RbLiXC6}
\end{table}

Examining the results of these computations:
\begin{itemize}
\item The $\svtotSC$ values for the JB$_2$ and JB$_1$ approximations are in good agreement with each other and with the FQMS $\svtot$ computations based on the KT and Medvedev potentials for the universal species  $^7$Li-X and $^{87}$Rb-X, where X = (N$_2$, Ar, Kr, and Xe). 
\item The $\svtotSC$ from JB$_0$ is systematically a larger than the other values for both $^7$Li-X and $^{87}$Rb-X in all cases, X = H$_2$, He, Ne, N$_2$, Ar, Kr, excepting $^{87}$Rb-Xe.
\end{itemize}
Thus, the $\svtotSC$ JB$_1$ values are a good heuristic to compare the universal species' $\svtot$ values against.

Next the $\svtotCsix$ results:
\begin{itemize}
\item There is consistency between the analytical expressions and their corresponding numerical computations of $\svtotCsix$ (a-JB$_0$~=~JB$_0$, and a-JB$_1$~=~JB$_1$). 
\item The JB$_0$ estimates for $\svtotCsix$ are systematically larger than the corresponding JB$_1$ and the JB$_2$ estimates, as one would expect since the JB$_0$ has the extra small $\hbar / \mu$ term in Eq.~\ref{eqB:svtotC6JB0}.
\item The $\svtotCsix$ JB$_1$ and JB$_2$ values agree well with each other. 
\end{itemize}
Thus, the $\eta_L(k, \rm{JB}_1)$ (Eq.~\ref{eqB:etaJB1}) are used for computing the $\svtotSC$ and $\svtotCsix$ in the main body of this paper.



\bibliography{Revising}
\end{document}